\documentclass{aa}  

\usepackage{xcolor}
\usepackage[colorlinks=true, citecolor=blue]{hyperref}
\usepackage{graphicx}
\usepackage{txfonts}

\begin{document}

   \title{Foreground removal in HI 21 cm intensity mapping under frequency-dependent beam distortions}


   \author{Athanasia Gkogkou\inst{1}
          \and
          Victor Bonjean\inst{1}
          \and
          Jean-Luc Starck\inst{2,1}
          \and
          Marta Spinelli\inst{3,4}
          \and
          Panagiotis Tsakalides\inst{5,6}
          }
           
   \institute{Institutes of Computer Science and Astrophysics, Foundation for Research and Technology Hellas (FORTH), Greece\\ \email{agkogkou@ia.forth.gr}
         \and
   Université Paris-Saclay, Université Paris Cité, CEA, CNRS, AIM, 91191, Gif-sur-Yvette, France
   \and
    Observatoire de la Côte d'Azur, Laboratoire Lagrange, Bd de l'Observatoire, CS 34229, F-06304 Nice Cedex 4, France
    \and
    Department of Physics and Astronomy, University of the Western Cape, Bellville, Cape Town 7535, South Africa
    \and
    Institute of Computer Science, Foundation for Research and Technology Hellas (FORTH), Greece
    \and
    Department of Computer Science, University of Crete, Greece
    }

   \date{Received 12/09/2025; Accepted 15/11/2025}

  \abstract
   {Neutral hydrogen (HI) intensity mapping with single-dish experiments is a powerful approach for probing cosmology in the post-reionization epoch. However, the presence of bright foregrounds over four orders of magnitude stronger than the HI signal makes its extraction highly challenging. While all methods perform well when assuming a Gaussian beam degraded to the worst resolution, most of them degrade significantly in the presence of a more realistic beam model.}
   {The complexity introduced by frequency-dependent beam effects motivates the need for methods that explicitly account for the instrument’s response. In this work, we investigate the performance of SDecGMCA. This method extends DecGMCA to spherical data, combining sparse component separation with beam deconvolution. Our goal is to evaluate this method in comparison with established foreground removal techniques, assessing its ability to recover the cosmological HI signal from single-dish intensity mapping observations under varying beam conditions.}
   {We use simulated HI signals and foregrounds informed by existing observational and theoretical models, covering the frequency ranges relevant to MeerKAT and SKA-Mid. The foreground removal techniques tested fall into two main categories: model-fitting methods (polynomial and parametric) and blind source separation methods (PCA, ICA, GMCA, and SDecGMCA). Their effectiveness is evaluated based on the recovery of the HI angular and frequency power spectra under progressively more realistic beam conditions.}
   {While all methods perform adequately under a uniform degraded beam, SDecGMCA remains robust when frequency-dependent beam distortions are introduced. In the oscillating beam case, SDecGMCA suppresses the spurious spectral peak at $k_\nu \sim 0.3$ and achieves $\lesssim 5\%$ accuracy at intermediate angular scales ($10 < \ell < 200$), outperforming other methods. Furthermore, masking bright Galactic regions significantly improves the recovery of the HI signal, particularly for SDecGMCA, which benefits most from excluding contaminated lines of sight. Beam inversion, however, remains intrinsically unstable beyond $\ell \sim 200$, setting a practical limit on the method.}
   {Our findings highlight the limitations of simple fitting and standard blind source separation (BSS) methods once realistic beam effects are considered, and establish SDecGMCA as a particularly promising approach for future single-dish intensity mapping surveys. Its robustness across beam models, combined with the improvements achievable through masking strategies and forthcoming refinements to its thresholding scheme, suggest that SDecGMCA could provide reliable spherical harmonics reconstructions of the HI power spectrum in upcoming experiments.}


   \maketitle
%

\section{Introduction}
\label{sect:introduction}

Achieving higher precision in probing the large-scale structure of the Universe with increasing precision requires observations over wider cosmic volumes. One of the most promising probes for this purpose is the 21 cm emission line of neutral hydrogen (HI), which traces the distribution of matter across cosmic time \citep{furlanetto2006, pritchard2012}. Unlike optical tracers, the 21 cm line is not obscured by dust and offers a tomographic view of the Universe, as the observed frequency directly maps to redshift.

Due to the faintness of the HI signal, detecting individual galaxies in 21 cm emission beyond the local Universe remains observationally expensive, requiring prohibitively long integration times \citep[e.g.,][]{fernandez2016, gogate2020}. HI intensity mapping \citep[IM,][]{battye2004, santos2005, chang2008, peterson2009, kovetz2017} circumvents this challenge by measuring the collective 21 cm emission from unresolved sources over large sky areas and broad frequency ranges \citep{battye2013, bull2015}. Although this technique sacrifices angular resolution, it enables fast surveys of cosmological volumes with excellent redshift precision.

Several experiments targeting the post-reionization epoch have made significant progress in detecting the HI signal through intensity mapping. The Green Bank Telescope (GBT) has produced cross-correlation measurements with galaxy surveys at low redshift \citep{masui2013, switzer2013, wolz2017, wolz2022}, while the Parkes radio telescope has reported similar results using data from HIPASS and other galaxy catalogs \citep{anderson2018, tramonte2020, li2021}. More recently, the MeerKAT radio telescope \citep{cunnington2023, carucci2024, meerklass_collaboration2025, chen2025} and the Canadian Hydrogen Intensity Mapping Experiment (CHIME) \citep{amiri2023, amiri2024} have added to the growing list of detections, primarily relying on cross-correlations or stacking techniques. Auto-correlation measurements of the HI power spectrum have also emerged, notably at $z \sim 0.32$ and $z \sim 0.44$ \citep{paul2023}. Complementary work has placed upper limits on the signal \citep[e.g.,][]{mazumder2025}. Achieving high-precision auto-correlation measurements of the cosmological HI power spectrum with well-controlled systematics remains a key goal for the field.

One of the main challenges in HI intensity mapping is contamination from astrophysical foregrounds. Synchrotron emission from the Milky Way, free–free radiation, and extragalactic point sources outshine the HI signal by up to five orders of magnitude, even at high Galactic latitudes \citep{santos2005,jelic2008}. These contaminants exhibit complex spatial and spectral structure, which already complicates separation from the cosmological signal.

In addition, instrumental systematics, in particular the frequency-dependent primary beam, further complicate the problem. The beam response couples to the foregrounds and introduces artificial spectral features that can mimic the 21\,cm signal \citep{matshawule2021}. The situation can be more severe when the beam shows non-smooth (e.g., oscillatory) frequency dependence, as measured for MeerKAT \citep{asad2021}. Other departures from the idealized chromatic response, such as those caused by instrumental optics or imperfect beam characterization, have also been shown to introduce systematic distortions in both simulations and real data \citep{shaw2015,chen2025}.

A variety of methods have been developed to address this challenge. Among the most widely used are Blind Source Separation (BSS) techniques, which aim to disentangle the HI signal from foreground contamination without relying heavily on prior models. Techniques such as Principal Component Analysis (PCA), Independent Component Analysis (ICA), FastICA, and Generalized Morphological Component Analysis (GMCA) have shown promising results in both theoretical and simulation-based studies \citep{liu2011, ansari2012, chapman2012, masui2013, switzer2013, wolz2014, bigot-sazy2015, alonso2015, shaw2015, olivari2016, zhang2016, wolz2017, carucci2020, soares2022}.

These BSS techniques, along with several model-fitting approaches, were systematically compared in \citet{spinelli2021}, where their performance was evaluated under various beam configurations. That study found that no method performs optimally across all regimes. Each technique exhibited specific limitations, including systematic offsets and artifacts arising from beam effects or intrinsic properties of the foreground and HI signals. 

With the rise of machine learning, there has been increasing interest in applying deep learning techniques to foreground removal and HI signal reconstruction \citep{makinen2021, ni2022, acharya2023, shi2024, ni2024, chen2024}. These approaches have demonstrated notable success in simulation-based studies, often leveraging large datasets such as CAMELS \citep{villaescusa2021}. However, their reliance on high-fidelity training simulations makes them sensitive to mismatches between training and real-world data. Incomplete knowledge of astrophysical foregrounds or instrumental systematics can bias these methods and potentially compromise their robustness. To avoid such dependencies, we opt for a fully blind, model-independent approach in this work.

A promising solution is to jointly address beam deconvolution and source separation, since the standard linear mixture model ($\mathbf{Y} = \mathbf{A} \mathbf{X}$) becomes invalid when the beam differs across frequency channels \citep{jiang2017}. In such cases, deconvolution and component separation must be performed simultaneously rather than sequentially. The DecGMCA method introduced in that work achieves this by incorporating the concept of sparsity, originally developed in GMCA \citep{bobin2007, starck2016}, while also addressing the deconvolution problem. Its extension to full-sky data, SDecGMCA \citep{carloni2021}, enables joint source separation and beam correction on the sphere. In this work, we apply SDecGMCA for the first time to simulated HI single-dish intensity mapping data at post-reionization frequencies. By revisiting the comparison of widely used foreground removal methods, we evaluate whether SDecGMCA can resolve key limitations reported in earlier analyses.

The paper is structured as follows. In Sect.\,\ref{sect:simulations}, we describe the simulations used to evaluate the performance of different foreground removal techniques. Sect.~\ref{sect:methods} describes the new method we propose along with the other component separation algorithms used for comparison. In Sect.\,\ref{sect:results}, we present and compare the performance of these methods across different instrumental configurations. In Sect.~\ref{sect:masking_effect}, we examine how a masking strategy can enhance the performance of the foreground removal methods. Sect.~\ref{sect:conclusions} places our findings in the context of previous work and provides a final summary.

\section{Simulations}
\label{sect:simulations}

In this section, we describe the simulated data used to test and validate the various foreground removal methods. To create a realistic representation of the sky in the frequency range $900\text{–}1400 , \rm MHz$, the simulations include multiple components: the 21 cm cosmological signal, galactic foregrounds such as synchrotron and free–free diffuse emission, and extragalactic foregrounds.

For each frequency and component, we generate HEALPix maps \citep{gorski2005} with $\rm N_{side} = 256$, corresponding to $\rm N_{pix} = 12 \times N_{\text{side}}^2$ pixels per map. These components are then combined into full-sky maps, convolved with various telescope beam models, and finally contaminated with white noise per frequency channel, calculated using standard thermal noise estimates.

\subsection{Astrophysical components}
\label{subsect:simulations_astro_components}

\subsubsection{HI cosmological signal}
At low redshifts, neutral hydrogen is expected to reside only in high-density regions where it is self-shielded from the ionizing effects of the cosmic ultraviolet background. As a result, HI traces the densest regions of the underlying dark matter field, making it a linearly biased tracer of dark matter. To simulate the HI distribution, we use the CRIME simulation\footnote{\url{http://intensitymapping.physics.ox.ac.uk/CRIME.html}} \citep{alonso2014}. This software approximates the dark matter field using a lognormal realization. In this work the adopted cosmological parameters are \{$\Omega_m, \Omega_{\Lambda}, \Omega_b, h$\} = \{0.3, 0.7, 0.049, 0.67\}. The initial simulation cube has a side length of $\rm 3 \, Gpc \, h^{-1}$ and is divided into $2048^3$ cells. Light-cone effects and redshift-space distortions are naturally incorporated into the simulation. The HI density field is then obtained by applying an assumed linear HI bias. Following the prescription in \citet{carucci2020}, we adopt a redshift-dependent linear HI bias of the form $b_{\rm HI}(z) = 0.3(1+z) + 0.6$ \citep{martin2012}, which is consistent with observational constraints at redshifts $z \lesssim 0.8$ \citep{switzer2013}. For the HI cosmic abundance, we assume a redshift evolution given by $\Omega_{\rm HI}(z) = 4 \times 10^{-4} (1+z)^{0.6}$, in agreement with measurements from \citet{crighton2015}.

\subsubsection{Synchrotron emission}

We adopt the synchrotron emission model implemented in the Python Sky Model 3 \citep[\texttt{PySM3};][]{zonca2021}, which describes the emission as a power law in frequency with a spatially varying spectral index. The synchrotron brightness temperature at frequency $\nu$ and sky position $p$ is given by:
\begin{equation}
T_{\rm sy}(\nu, p) = T_s(p) \left( \frac{\nu}{\nu_0} \right)^{\beta_{\rm sy} (p)},
\label{eq:temperature_sychrotron}
\end{equation}
where $T_s(p)$ is the amplitude map at the reference frequency $\nu_0$, and $\beta_{\rm sy}(p)$ is the spectral index map.

The amplitude template is based on the 408 MHz all-sky map reprocessed by \citet{remazeilles2015}, while the spectral index map is derived from a combination of the Haslam 408 MHz data and WMAP 23 GHz observations \citep{miville2008}. Small-scale fluctuations have been added directly to the intensity template to better emulate observed structure. This model corresponds to the “power-law” option of the Planck Sky Model v1.7.8 \citep{delabrouille2013}, updated with the Remazeilles version of the Haslam map.

\subsubsection{Free-free emission}

We also use the \texttt{PySM3} prescription for the free-free emission. The spatial template is based on the analytic model adopted in the Commander fit to the Planck 2015 data, following the formalism described by \cite{draine2011}. This template represents the free-free brightness temperature at 30 GHz on degree angular scales. To represent small-scale fluctuations, additional structure is added following the procedure outlined in \cite{thorne2017}. The native resolution of the map is inherited from \texttt{PySM2}.

The emission is scaled in frequency using a spatially uniform power-law spectral index of $\beta_{\rm ff} = -2.14$, yielding the following model for the free-free brightness temperature:
\begin{equation}
T_{\rm ff}(\nu, p) = T_{\rm ff}(p) \left( \frac{\nu}{\nu_0} \right)^{\beta_{\rm ff}},
\label{eq:temperature_ff}
\end{equation}
where $T_{\rm ff}(p)$ is the free-free amplitude map at the reference frequency $\nu_0 = 30,\mathrm{GHz}$.

\subsubsection{Extragalactic point sources}

For extragalactic radio sources, an empirical model is adopted \cite{battye2013}, which fits a fifth-order polynomial to observational source counts at $1.4 \, \rm GHz$ from various surveys. By integrating these source counts, we derive the mean temperature representing unresolved sources at $1.4 \, \rm GHz$.  The spatial distribution of these sources is characterized by the angular power spectrum, which consists of both clustering and Poisson components. The clustering term is derived from the observed angular correlation function of radio sources, while the Poisson component is obtained by integrating the differential source counts weighted by their flux squared. The total angular power spectrum is given by:  
\begin{equation}
    C_{\ell} = C_{\ell}^{\rm clust} + C_{\ell}^{\rm Poisson}.
\end{equation}
It is then transformed into pixel space, mapping the sources by using the HEALPix synfast routine. Point sources exceeding $0.01 \, \rm Jy$ are randomly injected as fully resolved sources, following \cite{olivari2016}:  
\begin{equation}
T_{\rm ps}(1.4 \, {\rm GHz}, p) = \left( \frac{\lambda^2}{2k_B} \right) \Omega_{\rm pixel}^{-1} \sum_{i=1}^{N}S_i,
\label{eq:temperature_point_sources_1.4GHz}
\end{equation}
where $\lambda^2/2k_B$ is the conversion factor between flux and brightness temperature, $\Omega_{\rm pixel}$ is the map pixel area, and $N$ represents the number of sources per steradian with flux $S$, as determined by the empirical fit of the differential counts. To scale this $1.4 \, \rm GHz$ estimate across the requested frequency range, we apply a power law with a spectral index $\beta_{\rm ps}(p)$ that varies across the sky. This index follows a Gaussian distribution centered at $-2.7$ with a standard deviation of 0.2 \citep{bigot-sazy2015}. Consequently, the temperature of extragalactic point sources at each pixel and frequency is given by:  
\begin{equation}
    T_{\rm ps}(\nu, p) = T_{\rm ps}(1.4 \, {\rm GHz}, p) \left( \frac{\nu}{\nu_0} \right)^{\beta_{\rm ps}(p)},
    \label{eq:temperature_point_sources}
\end{equation}
where $\beta_{\rm ps}(p) \sim \mathcal{N}(-2.7,0.2)$. Sources brighter than $1 \, \rm Jy$ are assumed to have been identified and removed from the data.\\

\subsection{Instrumental effects}
\label{subsect:simulations_instrumental}

After generating and merging all components, we account for two key instrumental effects in the maps. First, the application of the telescope beam, which smooths the signal. Second, the addition of uncorrelated thermal noise.

\subsubsection{Telescope beam models}
\label{subsubsect:tescope_beams}

Our goal is to evaluate the performance of foreground removal methods in the presence of a beam model, progressing from simple to more complex approaches. In this study, we focus on a single-dish experiment with specifications similar to the MeerKAT radio telescope, featuring a dish diameter of $13.5 \, \rm m$. However, our analysis remains broadly applicable to other instrument configurations. 

The most straightforward beam model assumes a spherically symmetric Gaussian profile with no side lobes. The beam response is given by:
\begin{equation}
    B(\theta) = e^{-\frac{\theta^2}{2 \sigma^2}},
    \label{eq:beam_response}
\end{equation}
where $\theta$ represents the angular distance from the beam center, and the standard deviation $\sigma$ is related to the full width at half maximum (FWHM) by:
\begin{equation}
    \sigma = \frac{\rm FWHM}{\sqrt{8 \, {\rm ln}2}}.
\end{equation}

The FWHM of the beam is frequency-dependent and is given by:
\begin{equation}
    {\rm FWHM} = \frac{\lambda}{D} = \frac{c}{\nu D},
    \label{eq:fwhm_D}
\end{equation}  
where $c$ is the speed of light, $\lambda$ the observing wavelength, $\nu$ the observing frequency, and $D$ the telescope dish diameter. To achieve a uniform angular resolution across all frequency channels, we degrade the maps to match the resolution of the lowest-frequency channel, corresponding to the largest ${\rm FWHM}$. This requires an additional smoothing step, where the standard deviation in Eq.~\ref{eq:beam_response} is adjusted using:
\begin{equation}
    \sigma_{\rm new} = \sqrt{\sigma_{\rm worst \, res}^2 - \sigma^2},
    \label{eq:new_sigma_degraded}
\end{equation}
where $\sigma_{\rm worst \, res} = {\rm FWHM_{max}} / \sqrt{8 \, {\rm ln}2}$ corresponds to the lowest-resolution channel, and $\sigma$ is the original beam standard deviation for a given frequency. 

When this additional smoothing step is applied, we refer to the resulting beam model as the "Gaussian degraded" model. Conversely, if no re-smoothing is performed and the beam varies naturally with frequency, we denote it as the "Gaussian evolving" beam model. These two models will be referenced throughout our analysis.

To achieve a more realistic approximation of the MeerKAT primary beam, we introduce an oscillatory term in Eq.\,\ref{eq:fwhm_D} to account for additional effects arising from the interaction between the primary and secondary reflectors, as observed and modeled by \cite{matshawule2021}. Their study compared the simple Gaussian beam assumption to a more accurate representation of the MeerKAT beam and characterized the frequency-dependent evolution of the FWHM using the following expression:  
\begin{equation}
    {\rm FWHM} = \frac{\lambda}{D} \left[ \sum_{d=0}^{8} a_d \nu^d + A {\rm sin} \left( \frac{2 \pi \nu}{T} \right) \right],
    \label{eq:FWHM_D_oscillating}
\end{equation}
where the first term is an 8th-degree polynomial \citep[with coefficients provided in Table 1 of][]{matshawule2021}, and the second term introduces a sinusoidal oscillation with a period of $T=20 \, \rm MHz$ and an amplitude of $A = 0.1 \, \rm arcmin$. Throughout this work, we refer to this refined beam model as the "Gaussian oscillating" beam model.

\subsubsection{Instrumental noise}
\label{subsubsect:instrumental_noise}

For this study, we adopt instrument and survey parameters representative of an SKA1-MID-like single-dish survey \citep{santos2015}. Using Eq.~\ref{eq:sigma_instrumental_noise}, we generate full-sky noise maps, treating the computed values as the per-pixel noise variance at each frequency channel.

The instrumental noise is modeled as Gaussian and uncorrelated, with uniform variance across the sky. Its standard deviation varies with frequency according to
\begin{equation}
    \sigma_{\rm N}(\nu) \;=\; T_{\rm sys}(\nu)\,
    \sqrt{\frac{4\pi f_{\rm sky}}{\Delta\nu\, t_{\rm obs}\, N_{\rm dishes}\, \Omega_{\rm beam}(\nu)}},
    \label{eq:sigma_instrumental_noise}
\end{equation}
where $T_{\rm sys}$ is the system temperature, $f_{\rm sky}$ the observed sky fraction, $\Delta\nu$ the channel width, $t_{\rm obs}$ the total observing time, $N_{\rm dishes}$ the number of telescope dishes, and $\Omega_{\rm beam}$ the beam solid angle. The latter is related to the full width at half maximum (FWHM) of the beam through $\Omega_{\rm beam}(\nu)=1.133\,\theta_{\rm FWHM}^2(\nu)$.

The system temperature, which includes contributions from both receiver noise and sky emission, is modeled as
\begin{equation}
    T_{\rm sys}(\nu) \;=\; T_{\rm instr} \;+\; 66 \left(\frac{300~\mathrm{MHz}}{\nu}\right)^{2.55} \ \mathrm{K},
    \label{eq:temperature_instrumental_noise}
\end{equation}
where $T_{\rm instr}$ is the instrumental contribution and $\nu$ is the observing frequency in MHz.

The adopted survey parameters are: observed sky fraction $f_{\rm sky}=1$ (full-sky coverage), channel width $\Delta\nu=1~\mathrm{MHz}$, total observing time $t_{\rm obs}=20{,}000~\mathrm{h}$, instrumental temperature $T_{\rm instr}=25~\mathrm{K}$, and number of dishes $N_{\rm dishes}=197$. The relatively long observing time was chosen to minimize noise impact and to focus the comparison on method performance.

\section{Methods}
\label{sect:methods}

An intensity mapping survey produces a map of the total brightness temperature $T$ for each frequency channel $\nu$. At any given sky pixel $p$, the observed temperature is a combination of three components: the cosmological HI 21\,cm signal ($T^{\mathrm{HI}}_{\nu,p}$), astrophysical foregrounds ($T^{\mathrm{cont}}_{\nu, p}$), and instrumental noise ($T^{\mathrm{noise}}_{\nu,p}$). This can be written as:
\begin{equation}
    T^{\mathrm{obs}}_{\nu,p} = T^{\mathrm{cont}}_{\nu, p} + T^{\mathrm{HI}}_{\nu,p} + T^{\mathrm{noise}}_{\nu,p}.
    \label{eq:cube_temperature}
\end{equation}

The objective is to extract the cosmological component $T^{\mathrm{HI}}_{\nu,p}$ from the total observed signal $T^{\mathrm{obs}}_{\nu,p}$. Since the HI signal is mixed with instrumental noise, the separation focuses on removing the dominant foreground component. This can be achieved through two main approaches: (i) fitting and subtracting a smooth continuum model, ideally isolating the residual HI+noise component or (ii) applying blind source separation (BSS) algorithms under the assumption that the continuum emission is sufficiently smooth to be represented by a limited number of components. In this section, we provide a detailed overview of both approaches: model-fitting methods and BSS techniques.

\subsection{Blind source separation (BSS)}
\label{subsect:bss_methods}

In many scientific problems, especially in astrophysics, the primary objective is to extract meaningful information from multichannel observations. BSS, an unsupervised matrix factorization technique, is widely employed for this purpose. Specifically, we consider data consisting of $N_c$ frequency channels, where each channel provides an observation ($\rm X_{\nu}$). This observation is a linear combination of $N_s$ underlying sources ($\rm S$), with the addition of some level of noise ($\rm N_{\nu}$). Therefore, for channel $\nu$, a single observation, which represents an image, can be described by the following linear mixture model:
\begin{equation}
    \rm X_{\nu} = \sum_{n=1}^{N_s} A_{\nu}^{n} S_{n} + N_{\nu},
    \label{eq:BSS_main_equation}
\end{equation}
where $A$ is the mixing matrix, which determines the contribution of each source $S_n$ to the observation at frequency $\nu$. The goal of any BSS method is to estimate both $A$ and $S$ in an unsupervised way, relying solely on the observed data $X$.

However, this is an ill-posed inverse problem, as multiple combinations of $A$ and $S$ can yield the same observations $X$. To overcome this ambiguity, additional prior information about the sources or the mixing matrix is required. Commonly used constraints include assuming statistical independence of the sources (ICA), enforcing non-negativity of both sources and mixing coefficients (non-negative matrix factorization), or leveraging source morphological diversity and sparsity in a specific representation domain (GMCA). The problem becomes even more complex when the data are blurred due to the instrument's point spread function (PSF) or beam. In such cases, beyond source separation, an additional deconvolution step is necessary to recover the true signal (SDecGMCA). In the following, we describe these different approaches in more detail.

\subsubsection{Principal component analysis (PCA)}
\label{subsubsect:PCA}

Principal component analysis \citep[PCA,][]{mackiewicz1993} is a statistical method assumes that the sources are uncorrelated and seeks to diagonalize the covariance matrix of the observations, given by $\mathbf{C} = \mathbf{X} \mathbf{X}^{\top}$. Achieving this means expressing the covariance matrix in the form:
\begin{equation}
    \mathbf{C} = \mathbf{U} \mathbf{D} \mathbf{U}^{\top},
    \label{eq:pca_covmat}
\end{equation}
where $\mathbf{D}$ is a diagonal matrix containing the eigenvalues. Since the observed data $\mathbf{X}$ is a linear combination of the source signals $\mathbf{S}$, weighted by the mixing matrix $\mathbf{A}$ ($\mathbf{X} = \mathbf{A} \mathbf{S}$), we can rewrite the covariance matrix as:
\begin{equation}
    \mathbf{C} = (\mathbf{A}\mathbf{S})(\mathbf{A}\mathbf{S})^{\top} = \mathbf{A} \mathbf{S} \mathbf{S}^{\top} \mathbf{A}^{\top} = \mathbf{U} \mathbf{D} \mathbf{U}^{\top},
\end{equation}
bringing it to the form of Eq.\,\ref{eq:pca_covmat}, where $\mathbf{D}=\mathbf{S} \mathbf{S}^{\top}$ and $\mathbf{U}=\mathbf{A}$. If PCA successfully diagonalizes $\mathbf{C}$, it effectively determines the mixing matrix $\mathbf{A}$, whose columns correspond to the principal components of the data. These principal components are the eigenvectors of $\mathbf{C}$, computed after centering $\mathbf{X}$ to have zero mean.

Because bright foregrounds dominate both the amplitude and variance of the observed signal, they are expected to be captured by the first few principal components, those corresponding to the largest eigenvalues. This property allows PCA to efficiently isolate and remove foreground contamination, revealing the underlying cosmological 21\,cm signal. As a result, PCA has become a widely adopted method for foreground cleaning in the HI intensity mapping community \citep[e.g.,][]{alonso2015,cunnington2021}.

\subsubsection{Fast independent component analysis (FastICA)}
\label{subsubsect:ICA}

This statistical method, which was developed in \cite{hyvarinen1999} estimates the mixing matrix $\mathbf{A}$ by leveraging the assumption that sources are statistically independent. FastICA maximizes statistical independence using non-Gaussianity as a proxy. According to the central limit theorem, the probability density function of a mixture of independent variables is more Gaussian than that of any individual variable. Therefore, by maximizing a statistical measure of non-Gaussianity, FastICA effectively isolates the statistically independent sources.

Before applying FastICA, the data must first be centered to zero mean. The method then achieves non-Gaussianity maximization by maximizing the negentropy of the sources, a measure that quantifies the distance of a distribution from Gaussianity, highlighting structured patterns within the sources. FastICA has been successfully applied to both simulated and observational HI intensity mapping data for foreground removal and cosmological signal recovery \citep[e.g.,][]{wolz2014,cunnington2019,hothi2021}.

\subsubsection{Generalized morphological component analysis (GMCA)}
\label{subsubsect:GMCA}

This method relies on two key assumptions. First, the sources are sparse, meaning they have mostly zero values in a suitable transformed domain, such as the wavelet domain. Second, the sources show morphological diversity. This means their non-zero values appear in distinct, non-overlapping regions. These assumptions are essential as they allow to dramatically improve the contrast between distinct components and eventually ease the separation process.

This method solves Eq.\,\ref{eq:BSS_main_equation} iteratively after first wavelet transforming the data from $\mathbf{X}$ to $\mathbf{X^{wt}}$, using the starlet wavelet dictionary \citep{starck2007}. GMCA aims to minimize the following cost function:
\begin{equation}
    \underset{\mathbf{A}, \mathbf{S^{wt}}}{\rm min} \, \sum_{i=1}^{N_s} \, \lambda_i ||\mathbf{S^{wt}i}||{1} + ||\mathbf{X^{wt}} - \mathbf{A} \mathbf{S^{wt}}||^2_{\rm F}, \label{eq:GMCA_cost_function}
\end{equation}
where the first term represents the sparsity constraint, and the second term is the data fidelity term. Here, $|| \cdot ||_1$ denotes the $\ell_1$ norm, defined as $|| \mathbf{M} ||_1 = \sum_{i,j} |m_{i,j}|$, and $||\cdot||_{\rm F}$ is the Frobenius norm, which is defined as $|| \mathbf{M} ||_{\rm F}^2 = \sum_{i} \sum_{j} |m_{i,j}|^2 = {\rm Trace}(\mathbf{M}\mathbf{M^{\top}})$. $\lambda_i$ are regularization coefficients, or sparsity thresholds, ensuring that noise is robustly removed. These coefficients are gradually decreased to a final level determined by the noise characteristics \citep{bobin2007}. GMCA\footnote{\url{https://www.cosmostat.org/statistical-methods/gmca}} is a blind method, meaning that no specific model is used for either $\mathbf{A}$ or $\mathbf{S}$. The only required inputs are the number of components, $N_s$, and the assumption of sparsity.

This method has been previously applied to cosmic microwave background (CMB) map reconstruction using WMAP and Planck data \citep{bobin2014,bobin2016}, to 21\,cm signal recovery \citep{chapman2013, carucci2020, marins2022}, and more recently to the study of supernova remnants in X-ray data \citep{godinaud2023}.

\subsection{Fitting methods}
\label{subsect:fitting_methods}

Alternative approaches involve fitting methods, which aim to smooth the signal, effectively filtering out noise and isolating the foregrounds that vary smoothly with frequency. Such methods have been widely used in the literature for foreground cleaning \citep[e.g.,][]{spinelli2021,fonseca2021}.

\subsubsection{Polynomial fitting}
\label{subsubsect:polfit}

The simplest approach to foreground subtraction involves modeling the foregrounds with a smooth frequency-dependent function. We represent this smooth baseline with a 6th-order polynomial for each line of sight ($p$) as a function of frequency $\nu$:
\begin{equation}
T_{\rm fg}(\nu, p) = \sum_{n=0}^{6} a_n(p) \, \nu^n.
\label{eq:polynomial_fit}
\end{equation}
Here, $a_n(p)$ are the polynomial coefficients for each line of sight, capturing the smoothly varying foreground component across frequencies. The polynomial fit is then subtracted from the total temperature maps, effectively flattening the signal and removing the smooth foreground baseline.

However, a second step is required to prevent overfitting bright HI sources, whose distinct profiles may be misinterpreted by the polynomial fit and subsequently removed. To address this, we perform sigma clipping after the initial polynomial subtraction setting the threshold to $5 \, \sigma$. This step identifies and masks outliers, effectively marking the locations of extremely bright sources. With these outliers excluded, we reapply the polynomial fit to capture only the smooth foreground baseline without being influenced by the bright sources.

\subsubsection{Parametric fitting}
\label{subsubsect:fit}

Another approach is to leverage prior empirical knowledge of the foregrounds about their spatial distribution and frequency evolution. Synchrotron, free-free emission, and extragalactic point sources exhibit strong degeneracy due to their similar spectral characteristics \citep{planck_collaboration2016}. However, we assume that at the frequencies of interest (MHz), the diffuse synchrotron and free-free emissions are the major contributors. Therefore, in this analysis, we only model these two components. We assume that the extragalactic contribution will be absorbed into one or both of these components. Consequently, for our parametric fit, we adopt the model:
\begin{equation}
    T(\nu) = \alpha_1(p) \, \left( \frac{\nu}{\nu_0} \right)^{-2.13} + \, \alpha_2(p) \,  \left( \frac{\nu}{\nu_0} \right)^{\beta(p)},
\label{eq:param_fit}
\end{equation}
where $\alpha_1(p)$, $\alpha_2(p)$, and $\beta(p)$ are free parameters determined for each line of sight $p$. The first term corresponds to the free-free emission, with the power law fixed at $-2.13$ \citep{bennett1992}. The second term accounts for the remaining diffuse emission, i.e., synchrotron, where the power law is treated as a free parameter.

\subsection{A new method for beam-aware foreground separation: SDecGMCA}
\label{subsubsect:SDecGMCA}

When observations are affected by frequency-dependent distortions, such as a varying telescope beam, standard BSS and model-fitting methods struggle to disentangle beam-induced effects from the intrinsic spectral and spatial properties of the sky components. To address this challenge, we apply SDecGMCA \citep[][hereafter C21]{carloni2021} for the first time in the context of HI intensity mapping. SDecGMCA extends DecGMCA by performing joint blind source separation and beam deconvolution on full-sky spherical data, incorporating the frequency-dependent beam response directly into the separation model.

In this framework, the standard BSS model is generalized as:
\begin{equation}
\mathbf{X}_\nu = (\mathbf{A}_\nu \mathbf{S}) \ast \mathbf{H}_\nu + \mathbf{N}_\nu,
\label{eq:SDecGMCA_main_equation}
\end{equation}
where $\mathbf{X}_\nu$ is the observed data at frequency $\nu$, $\mathbf{A}_\nu$ is the mixing matrix, $\mathbf{S}$ is the source matrix, $\mathbf{H}_\nu$ is the beam operator (assumed isotropic and thus diagonal in harmonic space), and $\ast$ denotes spherical convolution. $\mathbf{N}_\nu$ represents additive noise. The model is reformulated in the spherical harmonic domain as:
\begin{equation}
\hat{\mathbf{X}}_\nu^{\ell m} = \hat{{\rm H}}_\nu^\ell \mathbf{A}_\nu \hat{\mathbf{S}}^{\ell , m} + \hat{\mathbf{N}}_\nu^{\ell , m},
\label{eq:SDecGMCA_harmonic}
\end{equation}
where $\ell$ and $m$ are the spherical harmonic indices, and $\hat{{\rm H}}_\nu^\ell$ is the beam transfer function at frequency $\nu$. The beam affects only $\ell$ due to isotropy, simplifying the deconvolution. Throughout this work, boldface symbols (e.g., $\mathbf{X}$, $\mathbf{A}$, $\mathbf{S}$) denote matrices or linear operators, while non-bold symbols (e.g., $\hat{{\rm H}}_\nu^\ell$) represent scalar quantities. A hat (e.g., $\hat{\mathbf{S}}$) indicates that the quantity is expressed in the spherical harmonic domain.

The inverse problem of recovering $\mathbf{A}$ and $\mathbf{\hat{S}}$ is addressed by minimizing the following cost function:
\begin{equation}
\underset{\mathbf{A}, \hat{\mathbf{S}}}{\mathrm{min}} \ \left\| \boldsymbol{\Lambda} \odot \left( \hat{\mathbf{S}} \mathcal{F}^{\dagger} \mathbf{\Phi}^\top \right) \right\|_1 
+ \sum_{(\ell, m) \in \mathcal{D}} \left\| \hat{\mathbf{X}}^{\ell , m} - \mathrm{diag}(\hat{{\rm H}}^\ell) \mathbf{A} \hat{\mathbf{S}}^{\ell , m} \right\|_2^2,
\label{eq:SDecGMCA_cost_function}
\end{equation}
where $\mathbf{\Phi}$ is the starlet wavelet transform and $\mathcal{F}^{\dagger}$ is the inverse spherical harmonic transform. The first term promotes sparsity in the wavelet domain via soft-thresholding, with thresholds controlled by $\boldsymbol{\Lambda}$, while the second term enforces data fidelity in harmonic space.

\subsubsection*{Iterative algorithm and regularization strategy}

SDecGMCA proceeds in two main stages: a warm-up phase and a refinement phase. Each iteration alternates between estimating the source maps $\hat{\mathbf{S}}$ and the mixing matrix $\mathbf{A}$. The main steps and their corresponding hyperparameters are summarized below.

\begin{enumerate}
    \item[\textbf{(i)}] Initialization of $\mathbf{A}$: The mixing matrix is initialized via a PCA on data re-convolved to the lowest-resolution channel.
    
    \item[\textbf{(ii)}] Update of $\hat{\mathbf{S}}$: The sources are estimated using a Tikhonov-regularized least-squares solution for each $(\ell, m)$ (Eq.\,21 in C21), where the regularization coefficient $\epsilon_{n,\ell}$ is adaptively set according to the selected strategy (Eqs.\,24–28 in C21). The overall strength of the regularization is controlled by the hyperparameter $c$.
    
    \item[\textbf{(iii)}] Enforcing sparsity: 
    After the inverse harmonic transform, $\hat{\mathbf{S}}$ is projected onto an isotropic undecimated wavelet (starlet) basis with five scales. Sparsity is promoted via a soft-thresholding scheme, where the number of retained coefficients (active support) increases linearly from 0 to the hyperparameter $K_{\max}$ during the warm-up phase. During the refinement stage, an $\ell_1$ reweighting step is applied to mitigate thresholding bias.
    
    \item[\textbf{(iv)}] Update of $\mathbf{A}$: The mixing matrix is estimated via least squares over all spherical harmonics (Eq.\,14 in C21). After each update, the columns of $\mathbf{A}$ are normalized to prevent degeneracies, and negative entries are set to zero to enforce physical non-negativity (Eqs.\,15–16 in C21).
    
    \item[\textbf{(v)}] Refinement phase: 
    During refinement, the thresholding levels and regularization parameters are adapted more precisely (see Table\,1 in C21) to improve convergence and mitigate overfitting.
\end{enumerate}

Iterations continue until the relative change in the source estimate satisfies
$\| \hat{\mathbf{S}}^{(\mathrm{it})} - \hat{\mathbf{S}}^{(\mathrm{it-1})} \|_2\| / \hat{\mathbf{S}}^{(\mathrm{it})} \|_2 < \epsilon$, with $\epsilon = 10^{-2}$ during the warm-up phase and $\epsilon = 10^{-4}$ during refinement.

Among the regularization strategies proposed in C21, we adopt Strategy\,\#3 for the warm-up phase. This approach sets the Tikhonov regularization coefficient $\epsilon_{n,\ell}$ based on the inverse of the smallest eigenvalue of the matrix $\mathbf{M} = \mathbf{A}^{\top} \mathbf{\hat{H}}^{\ell,m} \mathbf{\hat{X}}^{\ell,m}$ (Eq.\,26 in C21). This formulation scales the regularization strength with harmonic mode, applying stronger regularization at higher multipoles (where noise dominates) while preserving large-scale features. For the refinement phase, we use Strategy\,\#4, which determines $\epsilon_{n,\ell}$ from the angular power spectra of the sources $\mathbf{S}$ and noise from the previous iteration, modulated by the hyperparameter $c$ (Eq.\,27 in C21).

In all applications presented in this work, the harmonic inversion is restricted to $\ell \leq 200$. The number of sources is set to $n_s = 3$, $4$, and $5$ for the degraded, evolving, and oscillating beam configurations, respectively, with increasing $n_s$ reflecting the rising complexity of the instrumental response. The SDecGMCA algorithm is described in detail in C21. For completeness and reproducibility, the main hyperparameter choices and implementation settings adopted in this work are summarized in Table~\ref{tab:sdecgmca_params}.

\begin{table}
    \centering
    \caption{Parameters adopted in this work.}
    \begin{tabular}{c|ccc}
        \hline
        Beam configuration & Degraded & Evolving & Oscillating \\
        \hline
        $n_s$  & 3 & 4 & 5 \\
        $c$ & $10^{-3}$ & $5 \times 10^{-3}$ & $10^{-2}$ \\
        $K_{\max}$ & 0.3 & 0.8 & 0.7 \\
        \hline
    \end{tabular}
    \label{tab:sdecgmca_params}
\end{table}

SDecGMCA has not previously been applied to HI intensity mapping. In this work, we evaluate its performance in this new context under realistic observational conditions, including frequency-dependent beam distortions and strong foreground contamination.

\section{Results}
\label{sect:results}

In this section, we apply SDecGMCA to the simulated data (Sect.~\ref{sect:simulations}) to assess its performance in reconstructing the cosmological HI 21\,cm signal, and compare it with the foreground removal methods described in Sect.~\ref{sect:methods}. We evaluate both the angular and frequency power spectra in order to capture fluctuations of the signal across the sky and along the line of sight. Additional quantitative diagnostics, including the correlation between the true and reconstructed HI maps and the variance of the fractional residual power, $\Delta C_{\ell}/C_{\ell}$, as a function of $\ell$, are presented in Appendix~\ref{ap:extra_diagnostics}.

\subsection{Data pre-processing: masking the peaks}
\label{subsect:HI_bright_masking}

Before applying any foreground-removal technique, we introduced a pre-processing step designed to improve foreground identification and, consequently, HI reconstruction. The procedure follows the logic of the polynomial fitting approach. First, we fitted a sixth-order polynomial along each line of sight to model the smooth baseline. Subtracting this fit flattened the spectra, after which we identified outliers via sigma clipping at a threshold of $5\sigma$ on the residuals. These outlier pixels, corresponding to bright HI peaks that are further amplified by instrumental noise, were masked and excluded from a second polynomial fit. The refit, performed on the remaining data, provides a less biased baseline estimate. Finally, the masked pixels were replaced by the value of this second-fit baseline, which effectively smooths the foregrounds and facilitates subsequent separation by both BSS and fitting methods. This pre-processing was applied uniformly to all input maps prior to component separation. Further implementation details are given in Appendix~\ref{ap:hi_bright_masking}. We find this step essential, as unmitigated peaks in the temperature profiles significantly degrade the performance of all tested methods.

\subsection{Results of foreground removal methods}
\label{subsect:actual_results}

\begin{figure*}[h!]
    \centering
    \begin{minipage}{0.33\textwidth}
        \centering
        \includegraphics[width=\textwidth]{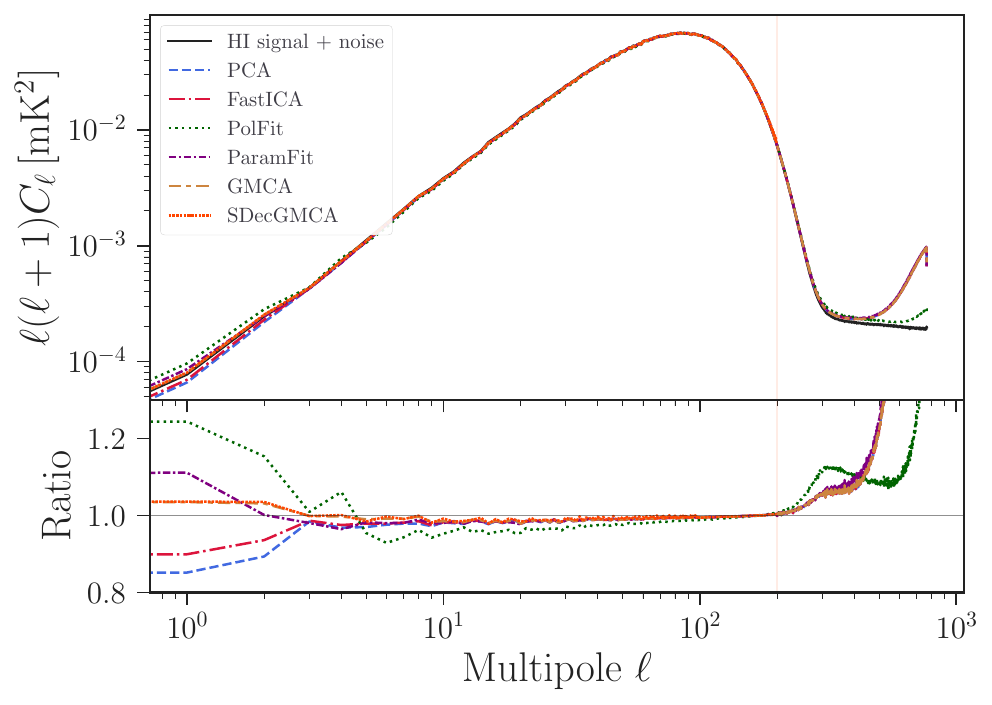}
    \end{minipage}
    \hfill
    \begin{minipage}{0.33\textwidth}
        \centering
        \includegraphics[width=\textwidth]{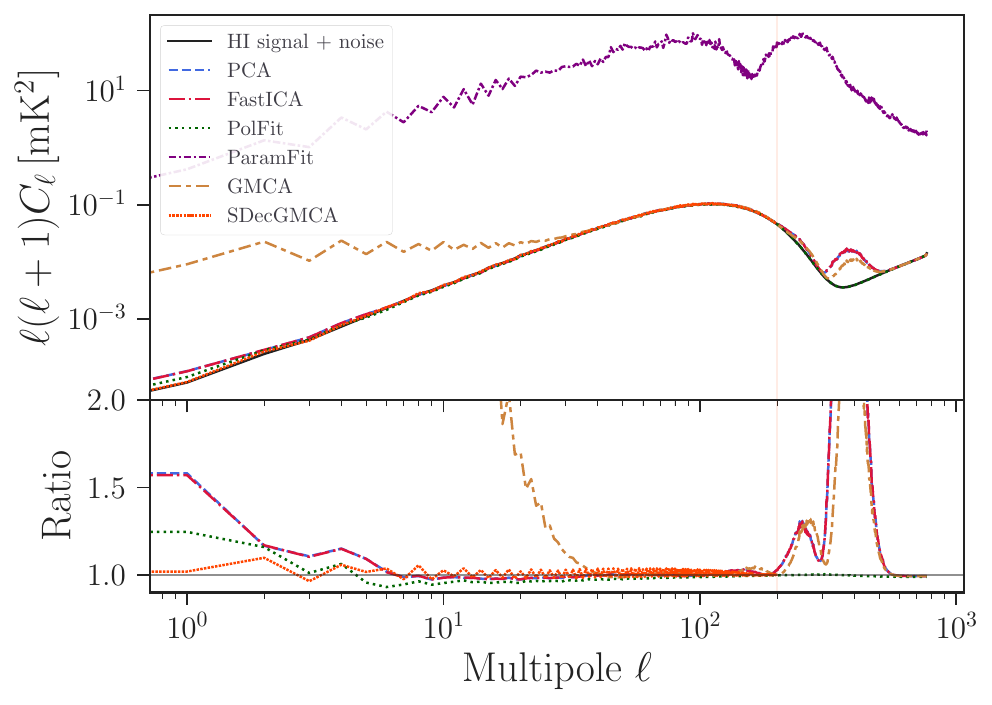}
    \end{minipage}
    \hfill
    \begin{minipage}{0.33\textwidth}
        \centering
        \includegraphics[width=\textwidth]{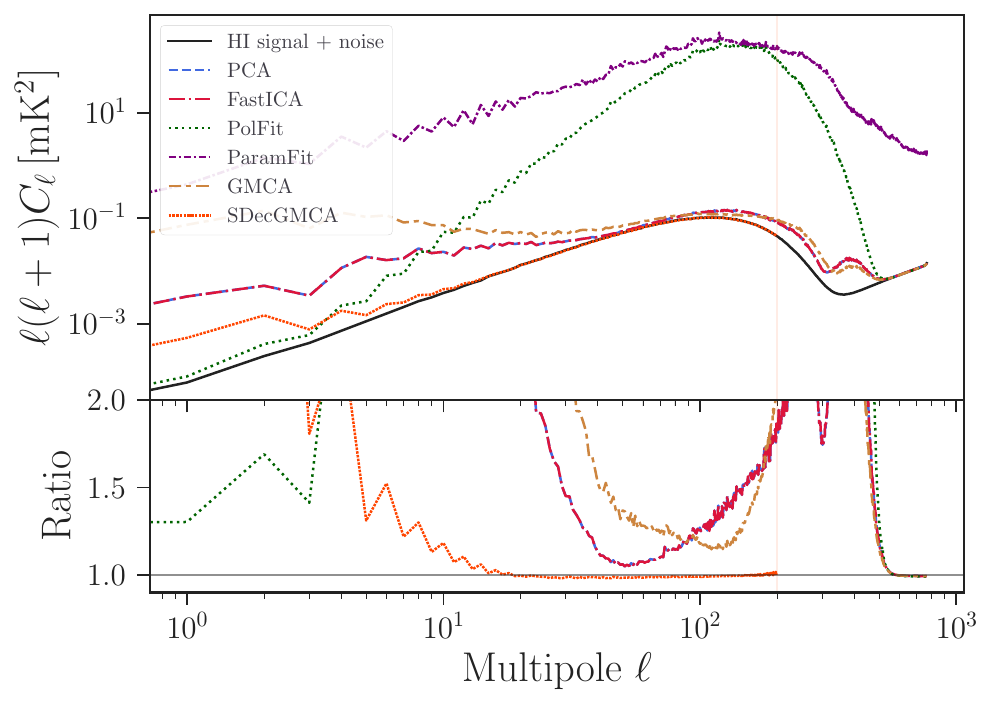}
    \end{minipage}
    \caption{Angular power spectrum of the reconstructed HI 21 cm signal using different methods, each represented by a distinct color. The bottom panel displays the ratio of the reconstructed to the true HI power spectrum for each method. From left to right, the panels correspond to different beam models: Gaussian degraded, Gaussian evolving, and Gaussian oscillating.}
    \label{fig:angular_powspec}
\end{figure*}

\begin{figure*}[h!]
    \centering
    \begin{minipage}{0.33\textwidth}
        \centering
        \includegraphics[width=\textwidth]{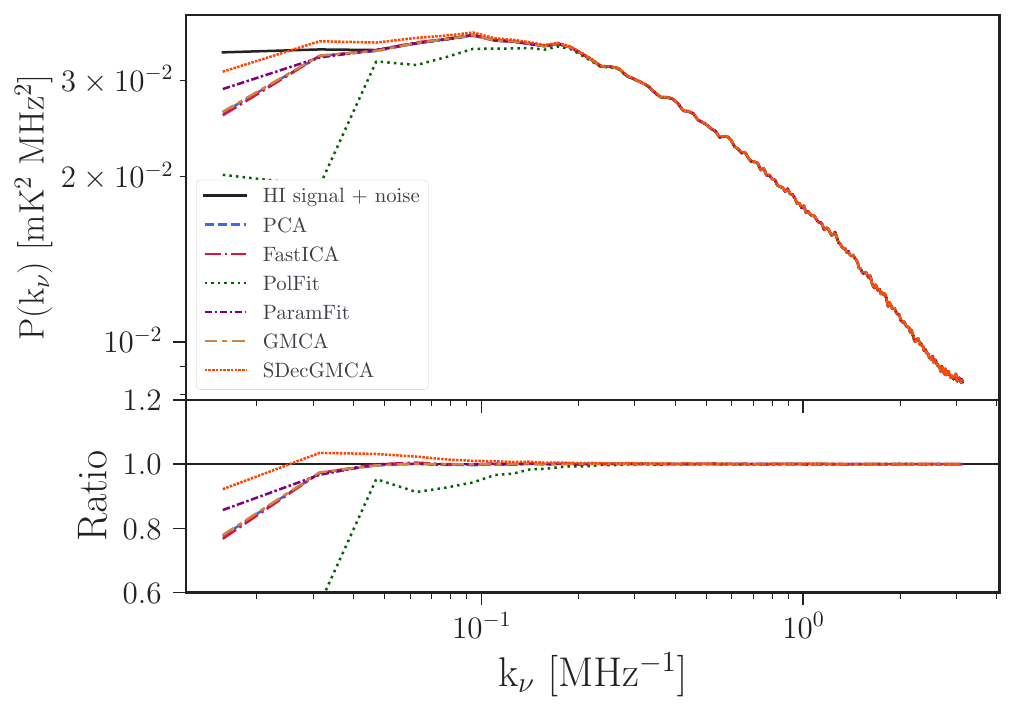}
    \end{minipage}
    \hfill
    \begin{minipage}{0.33\textwidth}
        \centering
        \includegraphics[width=\textwidth]{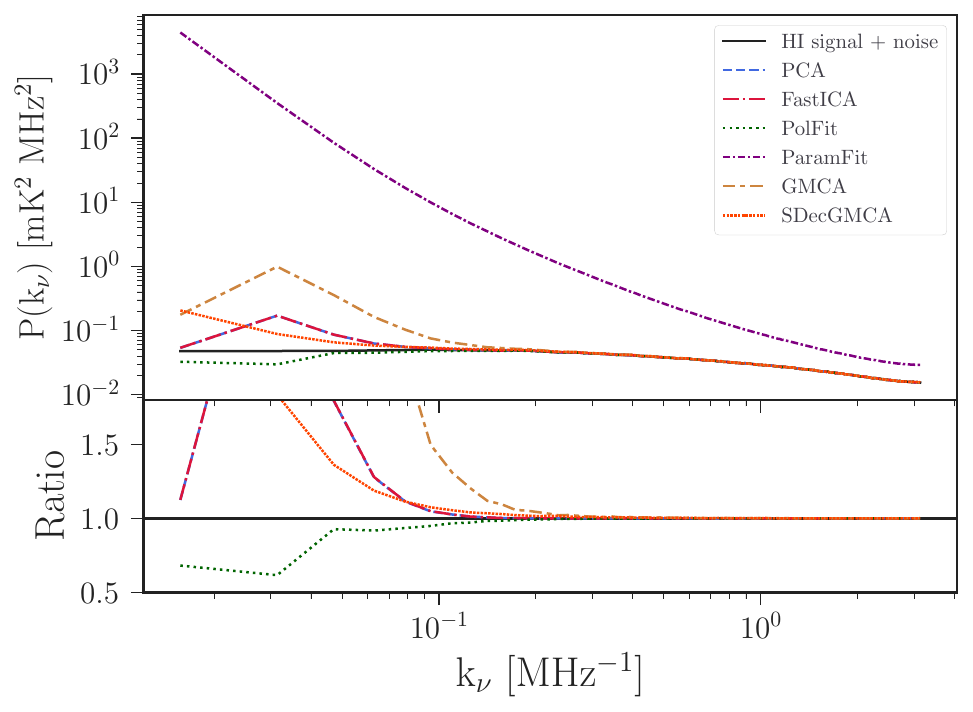}
    \end{minipage}
    \hfill
    \begin{minipage}{0.33\textwidth}
        \centering
        \includegraphics[width=\textwidth]{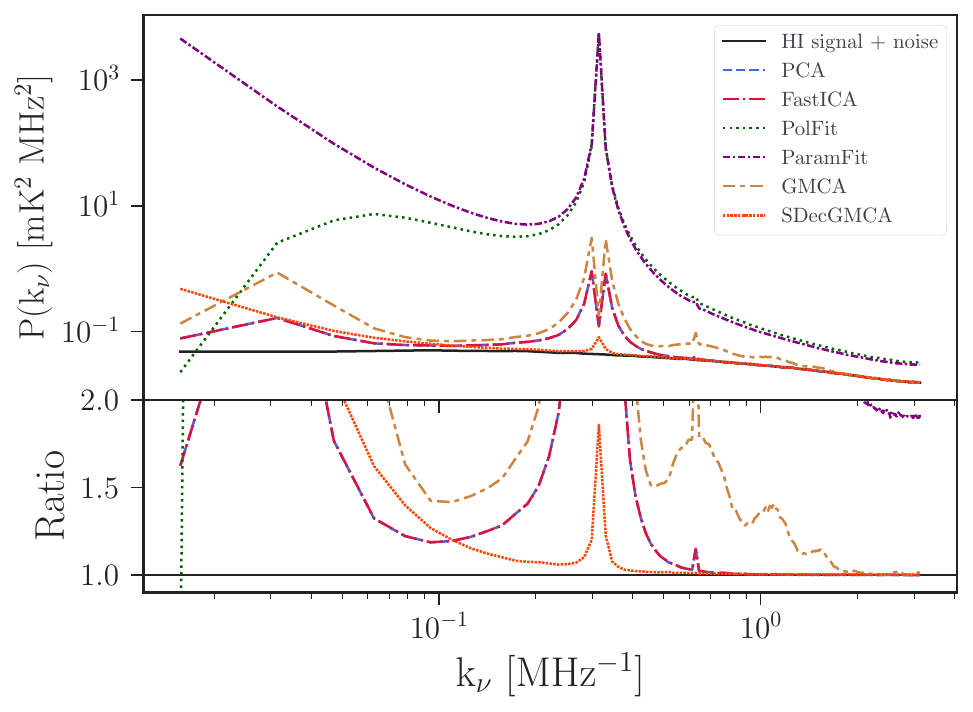}
    \end{minipage}
    \caption{Equivalent to Fig.\,\ref{fig:angular_powspec}, but showing the frequency power spectrum.}
    \label{fig:freq_powspec}
\end{figure*}

To investigate the impact of instrumental effects, we test three beam models (Sect.\,\ref{subsubsect:tescope_beams}): (i) a Gaussian-degraded beam with uniform resolution, (ii) a Gaussian-evolving beam with smooth frequency dependence, and (iii) a Gaussian-oscillating beam with oscillatory frequency dependence. Fig.\,\ref{fig:angular_powspec} and \ref{fig:freq_powspec} show the recovered power spectra for all methods across these beam scenarios.

Note that at high multipoles ($\ell > 200$), the reconstruction with SDecGMCA is limited by the implicit beam inversion that occurs when solving for $\hat{S}$ in the minimization problem (Eq.~21 of C21):
\begin{equation}
\rm \hat{S}_{\ell,m} \leftarrow 
\Big( A^{\top} \mathrm{diag}(H^{\ell})^2 A + \mathrm{diag}(\epsilon_{n_s,\ell}) \Big)^{-1} 
A^\top \, \mathrm{diag}(\hat{H}_\ell) \, \hat{X}_{\ell,m}.
\label{eq:sdecgmca_update}
\end{equation}
This inversion is stabilized by a Tikhonov regularization controlled by the hyperparameter $\epsilon_{n_s,\ell}$. While the regularization smooths the foregrounds at scales where the beam response is poor, it also introduces a negative bias in the foreground estimate, which translates into a spurious positive bias in the recovered 21\,cm signal. As a result, the regularization itself becomes a source of systematic error. In practice, the deconvolution can only be performed up to the multipole range where the matrix $\rm A^{\top} \mathrm{diag}(H^{\ell})^2 A$ remains invertible without regularization (i.e. $\epsilon_{n_s,\ell}=0$). This defines a maximum multipole $\ell_{\rm max}$, which depends on the beam width, beyond which the inversion becomes unstable. In our simulations this corresponds to $\ell_{\rm max} \simeq 200$, and indeed a prominent spurious peak appears across all beam models for $\ell>200$. For this reason, we cut in our plots the SDecGMCA-reconstructed angular power spectrum at $\ell=200$.

\subsubsection{Gaussian degraded beam}

In the simplest case of a Gaussian degraded beam, all methods perform well, recovering both the angular and frequency power spectra with better than 10\% accuracy across a wide range of scales. In this case, we used $n_s = 3$ for all the BSS methods. In the angular domain, the reconstruction remains within 10\% accuracy for $3 < \ell < 200$, while in the frequency domain all methods achieve similar accuracy for $k_{\nu} > 3 \times 10^{-2}$. At larger frequency scales ($k_{\nu} < 3 \times 10^{-2}$), a small underfitting is observed, reflecting the degeneracy between the long-wavelength foreground fluctuations and the cosmological HI signal, an ambiguity that none of the methods can resolve. Overall, such good performance is expected, as the uniform angular resolution preserves the spectral smoothness of the foregrounds across frequency channels, which facilitates their separation from the cosmological signal. In this case, we used $n_s = 3$ for all the BSS methods.

\subsubsection{Gaussian evolving beam}

When the beam evolves with frequency, the performance of the methods begins to diverge. In this case, we used $n_s = 4$ for all the BSS methods in order to accommodate the extra spectral features introduced by the beam evolution. 

The parametric fitting approach fails entirely, overestimating the angular power spectrum by several orders of magnitude across all scales. This likely occurs because the model, which fits only three free parameters per line of sight, lacks the flexibility to capture the additional spectral distortions induced by the frequency-dependent beam. In contrast, the BSS methods recover the angular power spectrum to within approximately 10\% accuracy at $4 \lesssim \ell \lesssim 200$, with the exception of GMCA, which shows a clear excess at intermediate and small multipoles. This degradation is likely related to GMCA’s reliance on morphological sparsity, which becomes less effective once the evolving beam disrupts the spatial coherence of the foregrounds across frequencies. 

In the frequency domain, all methods except the parametric fit successfully recover the power spectrum down to $k_{\nu} \sim 0.1$, below which deviations increase. At these larger frequency scales (lower $k_{\nu}$), the beam evolution couples angular and spectral modes, producing scale-dependent distortions that the BSS methods cannot fully disentangle. The polynomial fitting method, by construction, is insensitive to this particular effect. However, it still suffers from the intrinsic degeneracy between the smooth frequency dependence of the foregrounds and the large-scale fluctuations of the cosmological signal.

\subsubsection{Gaussian oscillating beam}

The most realistic and challenging case is that of the oscillating beam, which further complicates the separation of the cosmological signal from the foregrounds. In this scenario, we increased the number of sources to $n_s = 5$ in order to better capture the enhanced complexity introduced by the beam oscillations. The two fitting approaches (parametric and polynomial) fail entirely, yielding highly biased estimates of both the angular and frequency power spectra. The BSS methods perform comparatively better, although significant limitations remain. For this reason, in the following we focus exclusively on the performance of the BSS methods.

In the angular power spectrum, a strong excess appears at $\ell < 10$, similar to the offset observed in the previous cases but now significantly larger. This excess may arise from artificial residuals introduced during the pre-processing step that masks bright HI sources (Sect.~\ref{subsect:HI_bright_masking}). In the oscillating beam case, the initial polynomial subtraction may leave residual oscillations that are then amplified. During the subsequent sigma-clipping and re-fitting stage, these residuals can be misidentified as bright HI sources, leading to incorrect replacement of foreground-dominated pixels with smoothed fits. Combined with the frequency-dependent beam evolution, these misattributions can generate large-scale artificial structures in the maps. Since they are partially interpreted as HI signal, they produce an excess at large angular scales in both the angular and frequency spectra. At intermediate scales ($10 < \ell < 200$), SDecGMCA stands out as the only method capable of recovering the HI signal with an accuracy better than 5\%, whereas PCA, FastICA, and GMCA show noticeably larger residuals.

In the frequency domain, all methods display a spurious peak around $k_{\nu} \sim 0.3$. A similar artifact was also reported by \citet{matshawule2021} and \citet{spinelli2021}, where it was attributed to beam–foreground coupling. Among the methods tested here, SDecGMCA is the most effective at suppressing this peak, underscoring the importance of accounting for the beam within the separation process. Nevertheless, an excess of power persists at low $k_{\nu}$ across all BSS methods, and in this respect SDecGMCA performs no better than PCA or FastICA.

These results highlight the limitations of both standard BSS techniques and simple fitting approaches when confronted with complex, frequency-dependent instrumental effects. In line with \citet{spinelli2021} and \citet{matshawule2021}, we find that evolving and oscillating beams severely degrade the performance of traditional methods, producing artifacts in both angular and frequency domains. By contrast, SDecGMCA, which jointly performs source separation and beam deconvolution, substantially mitigates some of these effects, providing the most consistent recovery of the HI signal across all beam scenarios. Nevertheless, the intrinsic instability of beam inversion, already well known in the literature, remains a major limitation. Overall, these findings establish SDecGMCA as a particularly promising method for realistic 21\,cm intensity mapping experiments. It is worth noting that in our analysis we applied the method to the full sky without removing or masking any regions, and even under these conditions SDecGMCA shows encouraging performance. As we discuss in Sect.~\ref{sect:masking_effect}, a targeted masking strategy of the most contaminated lines of sight can further mitigate residual artifacts in the reconstructed maps.

\subsection{Dependence on the number of components}
\label{subsect:ns_dependency}

\begin{figure}
    \centering
    \includegraphics[width=\linewidth]{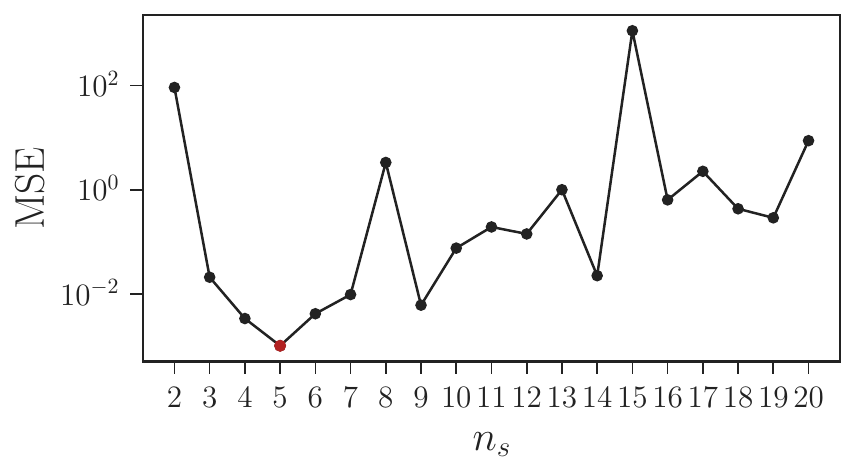}
    \caption{Total MSE of the SDecGMCA reconstruction as a function of the number of sources $n_s$. The MSE is computed with respect to the ground truth by combining both the frequency and angular power spectra. The curve highlights the trade-off between underfitting at low $n_s$ and overfitting at high $n_s$, with the minimum occurring around $n_s = 5$ (marked in red), which provides the most balanced reconstruction.}
    \label{fig:MSE_ns}
\end{figure}

The performance of SDecGMCA, as with other BSS methods, depends not only on the instrumental complexity but also on the internal choices made within the algorithm. One of the most critical hyperparameters is the number of sources ($n_s$) assumed in the separation model. Since $n_s$ sets the dimensionality of the component space, its value strongly influences the balance between foreground suppression and signal recovery. To assess this dependence, we applied SDecGMCA to the oscillating beam simulations, varying $n_s$ between 2 and 20, and for each run computed the mean squared error (MSE) between the reconstructed and true HI signal in both the frequency and angular power spectra (Fig.~\ref{fig:MSE_ns}).

\begin{figure*}
    \centering
    \includegraphics[width=1\linewidth]{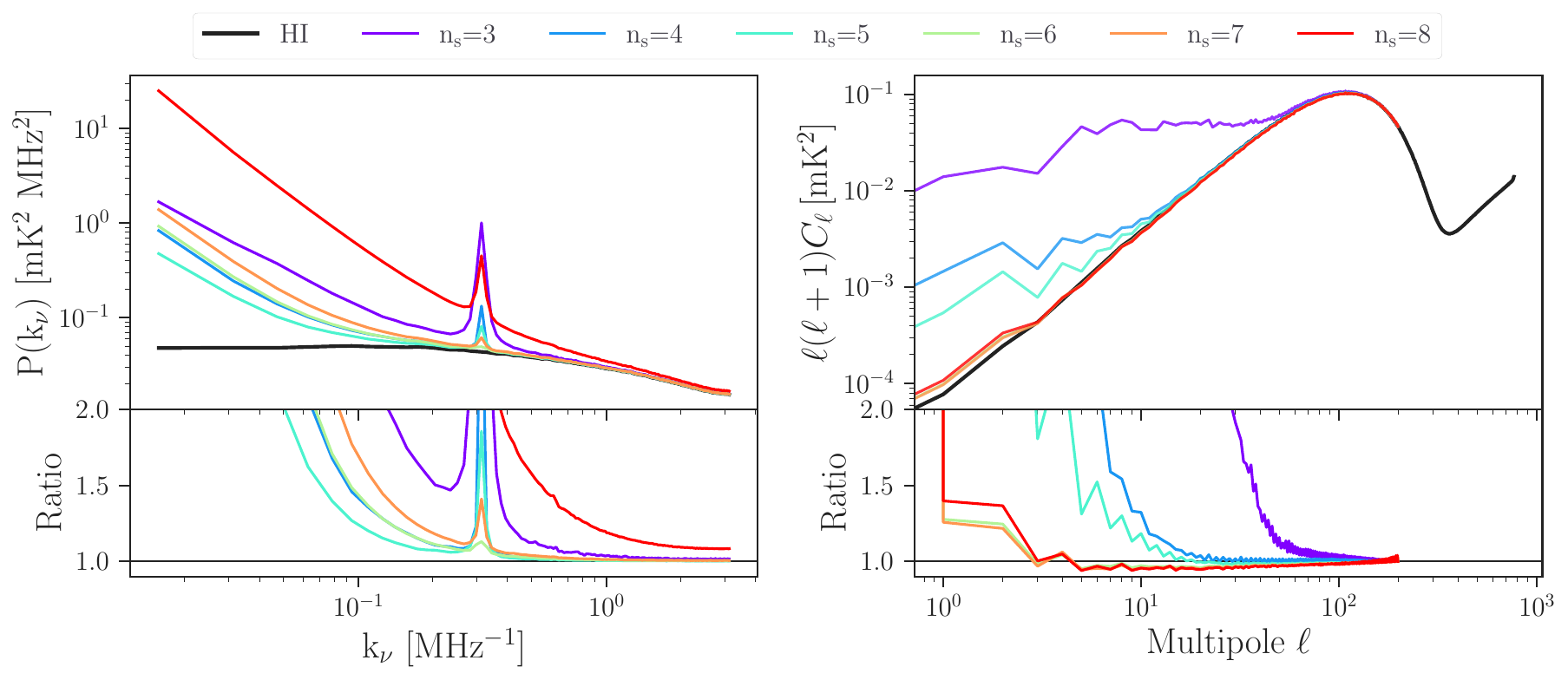}
    \caption{Reconstructed HI frequency (left) and angular (right) power spectra obtained with SDecGMCA for different choices of the number of sources ($n_s$, shown in different colors). The black line shows the input HI signal for reference, and the lower panels display the ratio between the reconstruction and the true signal.}
    \label{fig:mult_ns}
\end{figure*}

A closer inspection of the reconstructed spectra shows (Fig.\,\ref{fig:mult_ns}) that the optimal choice of $n_s$ is somewhat scale-dependent. At large frequency scales (low $k_\nu$), $n_s=5$ provides the best suppression of excess power, while the localized bump around $k_\nu \sim 0.3$ is more efficiently removed when $n_s=6$ is used. In the angular domain, differences between models are less pronounced at very large multipoles, where beam effects dominate, but at intermediate scales ($\ell \sim 10^1$--$10^2$) the reconstruction improves noticeably for $n_s=7$. This behavior reflects the fact that different astrophysical foregrounds and instrumental systematics dominate different scales, requiring varying degrees of model flexibility.

Overall, the most balanced reconstruction across both domains is obtained for $n_s = 5$. As Fig.~\ref{fig:MSE_ns} illustrates, the number of sources plays a role analogous to regularization. Too few sources (e.g. $n_s=2$--3) lack the flexibility needed to capture the complex spectral behavior of the foregrounds, causing residual contamination to leak into the HI signal. Too many (e.g. $n_s \gtrsim 10$) risk overfitting, as the algorithm tends to fragment coherent astrophysical components into multiple modes, reintroducing noise and residuals. This trade-off is clearly visible in the MSE curve, which rises again at high $n_s$. The importance of carefully choosing the number of sources has also been emphasized in the HI extraction literature, where a range of BSS methods applied to different experimental configurations consistently highlight this trade-off \cite[e.g.,][]{chapman2013,mertens2018,carucci2020,hothi2021,carucci2024}.

In summary, while SDecGMCA is capable of robustly recovering the HI power spectra, its performance remains sensitive to the choice of $n_s$. A moderate number of sources ($n_s \simeq 5$--7) provides the best compromise across scales, although the precise optimum can vary depending on the $k_\nu$ or $\ell$ range of interest. A practical challenge arises when applying the method to real data, where the ground truth is not available and the optimal value of $n_s$ cannot be identified directly. In such cases, indirect diagnostics such as the eigenvalues of the covariance matrix of the signal, the stability of the recovered power spectra, or cross-correlations with external tracers will be needed.

Another important set of hyperparameters in SDecGMCA are those associated with the thresholding procedure. We have not discussed them here in detail, as they are more technical and beyond the scope of this work. In a follow-up study, however, we will show that replacing the current thresholding scheme with a structured convolutional neural network that acts as a learned, wavelet-like multiscale representation \citep{bonjean2025} can effectively eliminate the dependence of the method on these hyperparameters.

\section{Masking effect}
\label{sect:masking_effect}

Masking the bright peaks in the frequency profiles of individual lines of sight (Appendix\,\ref{ap:hi_bright_masking}) helps improve the HI reconstruction. These peaks correspond to bright HI sources and not to extragalactic point sources, which has already been studied in \citet{matshawule2021}. In particular, this masking of the peaks mitigates systematic offsets in the frequency power spectrum, such as the broadband bias across all scales reported in \citet{spinelli2021} and also observed in our results. While SDecGMCA demonstrates improved performance under these conditions, particularly in the challenging oscillating beam case, there remain regions of the sky where the method fails to accurately reconstruct the underlying HI signal. A likely contributing factor is the inclusion of the Galactic plane and other extremely bright lines of sight in the input maps, which make the separation even more challenging.

In this section, we examine how excluding or attenuating these bright-foreground regions influences the quality of the HI signal reconstruction. Although directly applying a Galactic mask during component separation is currently incompatible with the SDecGMCA framework, we explore two indirect approaches: first, by post-processing the reconstructed HI signal to exclude lines of sight with problematic reconstruction (Sect.\,\ref{subsect:post_masking}) and second, by artificially dimming the foreground amplitude in and around the Galactic plane to simulate the effect of a Galactic mask (Sect.\,\ref{subsect:dimmed_frg}).

\subsection{Post-masking}
\label{subsect:post_masking}

To investigate the origin of the spurious peak in the frequency power spectrum and the excess power at low $k_{\nu}$, we closely examined the reconstructed HI maps from all foreground removal methods. We observed that, for a subset of lines of sight, there is coherent spectral structure as a function of frequency, indicative of leaked foregrounds. To identify which lines of sight contribute to the residuals, we computed the MSE between the reconstructed and true HI signal along each line of sight ($\rm MSE_{los}$). We then flagged the lines of sight with $\lvert \langle \mathrm{MSE} \rangle - \mathrm{MSE_{los}} \rvert > \sigma_{\mathrm{MSE}}$, where $\langle \cdot \rangle$ denotes the mean and $\sigma$ the standard deviation over all lines of sight. 

\begin{figure*}[h!]
    \centering
    \includegraphics[width=0.9\linewidth]{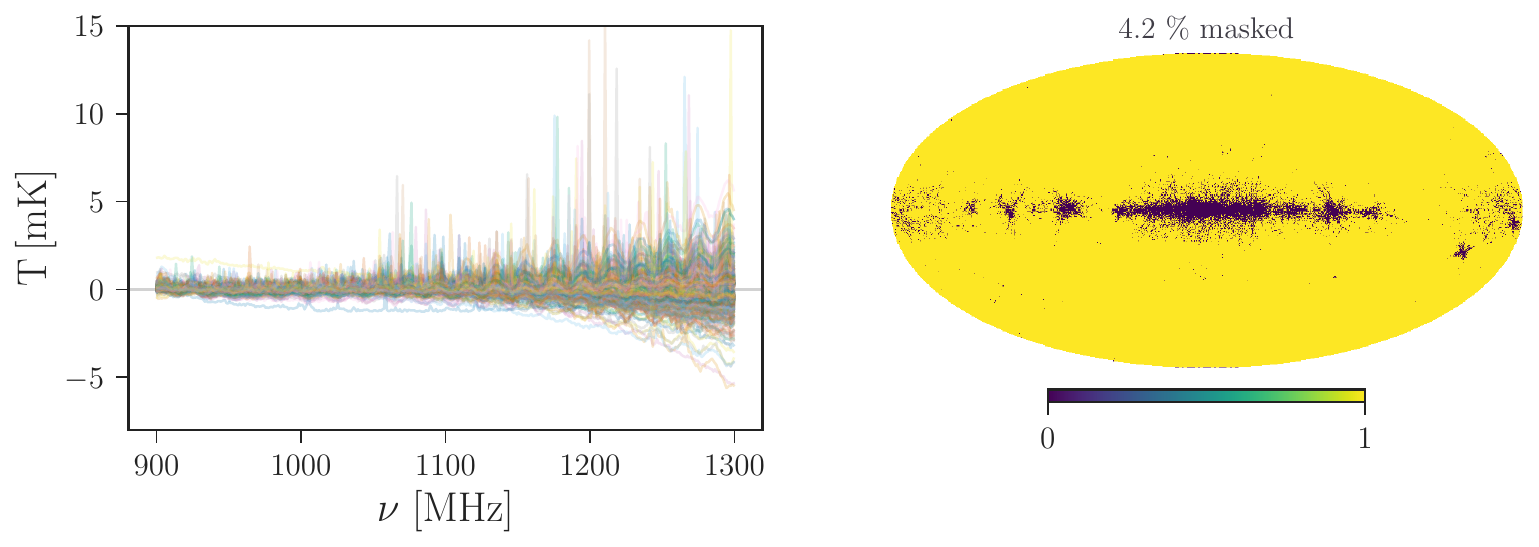}
    \caption{Filtering lines of sight prior to computing the power spectrum of the reconstructed HI signal. Left: Brightness temperature of the reconstructed HI signal as a function of frequency for all flagged lines of sight that deviate from the expected spectral flatness. Right: Sky map showing the location of the flagged lines of sight, illustrating the regions excluded by the constructed mask. The percentage of lines of sight that are masked is indicated at the top of the map.}
    \label{fig:filtered_lines_of_sight}
\end{figure*}

\begin{figure*}[h!]
    \centering
    \begin{minipage}{0.48\textwidth}
        \centering
        \includegraphics[width=\textwidth]{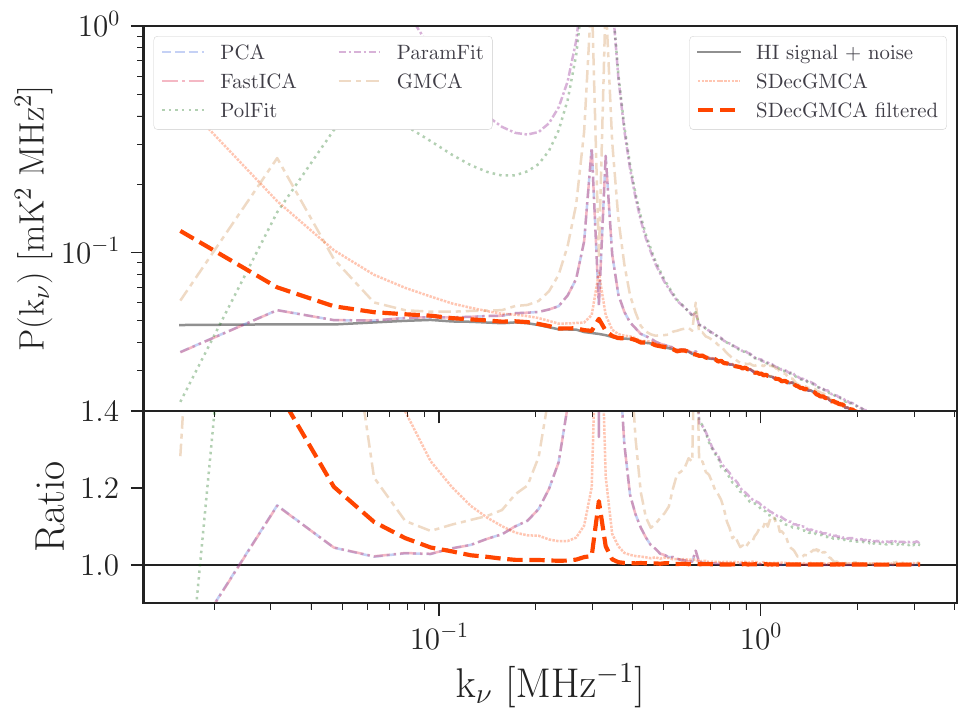}
    \end{minipage}
    \hfill
    \begin{minipage}{0.5\textwidth}
        \centering
        \includegraphics[width=\textwidth]{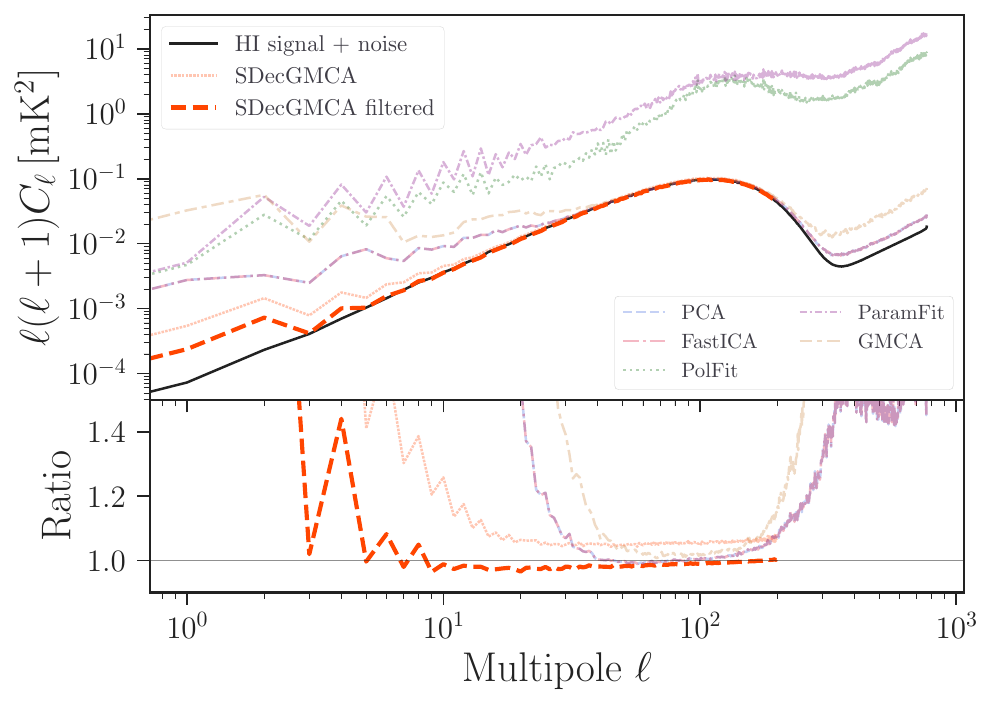}
    \end{minipage}
    \caption{Impact of post-processing masking on the HI signal reconstruction for the oscillating beam case. Left: Frequency power spectrum of the reconstructed HI signal. The solid black curve shows the true HI signal plus noise, while the dotted red curve represents the SDecGMCA reconstruction before masking. The dashed red curve shows the result after excluding problematic lines of sight, which suppresses the spurious peak and improves recovery at lower $k_{\nu}$. Other methods are shown in lighter transparent lines for comparison, with minimal improvement observed. Right: Same as the left panel, but for the angular power spectrum of the reconstructed HI signal. The top panel displays the reconstructed power spectra from different methods, while the bottom panel shows the ratio to the true signal.}
    \label{fig:postprocess_mask_freq_ang_powepsec}
\end{figure*}

The resulting flagged lines of sight are illustrated in Fig.\,\ref{fig:filtered_lines_of_sight}, both as spectral profiles (brightness temperature versus frequency) and spatially on the sky. Visually, it is evident that these contaminated lines of sight are located along the Galactic plane, where foreground emission is particularly strong. To evaluate their influence on the reconstructed power spectra, we excluded the flagged regions when computing the frequency and angular power spectra of the recovered HI maps. The results of this analysis are presented in Fig.\,\ref{fig:postprocess_mask_freq_ang_powepsec}.

Applying this post-processing reduces the amplitude of the peak in the frequency power spectrum further, as shown in the left panel of Fig.\,\ref{fig:postprocess_mask_freq_ang_powepsec}. Moreover, the HI signal is recovered down to lower $k_{\nu}$ scales, significantly improving the fidelity of the reconstruction. To ensure a fair comparison, we applied the same masking procedure to the outputs of the other methods. However, no significant improvement was observed in these cases. This suggests that their performance degradation is introduced during foreground removal and exceeds what can be mitigated by post-processing.

In addition, we applied the mask described in the previous subsection to the angular power spectrum. Masking the Galactic plane before spectrum estimation is a standard practice in the CMB literature \citep[e.g.,][]{bennett2003,planck_collaboration2016}, where regions of bright Galactic emission are excluded to reduce contamination and improve the robustness of the cosmological inference. The resulting angular power spectra, shown in the right panel of Fig.~\ref{fig:postprocess_mask_freq_ang_powepsec}, demonstrate the clear benefits of this strategy. Excluding the Galactic plane significantly improves the reconstruction of the HI signal for all BSS methods. Among them, SDecGMCA benefits the most, successfully recovering the HI power spectrum down to smaller multipoles compared to the case without post-processing masking.

\subsection{Simulating a Galactic mask}
\label{subsect:dimmed_frg}

Encouraged by the improvements observed when excluding bright regions from the power spectrum computation, we next explore whether incorporating a masking strategy before applying the separation methods could yield similar or even stronger benefits. Since direct masking is not currently supported by SDecGMCA, we simulate this scenario by artificially dimming the foregrounds in and around the Galactic plane. This approach mimics an observational strategy that avoids highly contaminated regions of the sky and allows us to evaluate how bright Galactic structures impact the reconstruction quality.

\begin{figure*}[h!]
    \centering
    \begin{minipage}{0.48\textwidth}
        \centering
        \includegraphics[width=\textwidth]{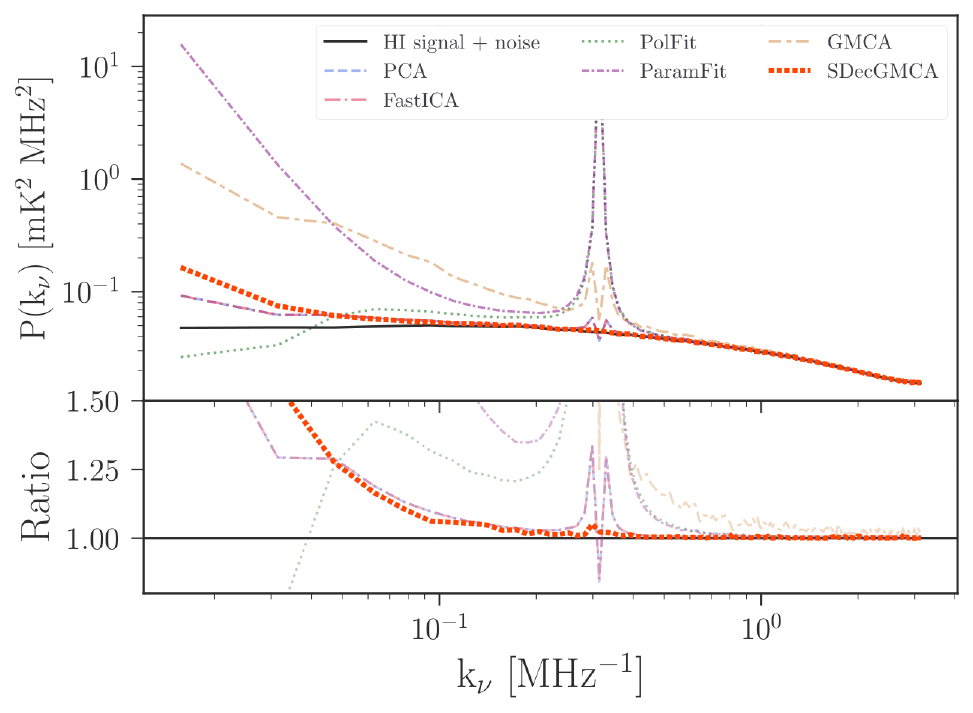}
    \end{minipage}
    \hfill
    \begin{minipage}{0.5\textwidth}
        \centering
        \includegraphics[width=\textwidth]{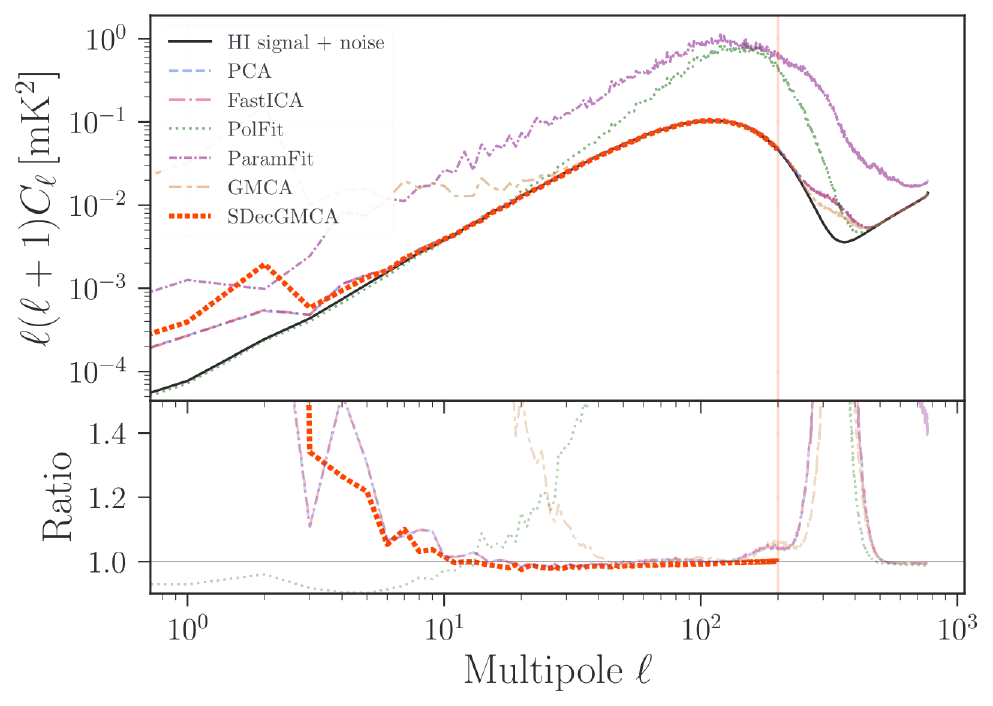}
    \end{minipage}
    \caption{Frequency (left) and angular (right) power spectra of the reconstructed HI signal obtained with different foreground removal methods, shown in various colors. All methods were applied to simulated maps with artificially dimmed foregrounds. These maps were convolved with a Gaussian oscillating beam prior to signal reconstruction.}
    \label{fig:dimmed_results}
\end{figure*}

To test this idea in a controlled and continuous way, we applied a smooth Gaussian weighting function in Galactic latitude to the simulated foreground maps. The weights preserve the signal outside $|b| > 15^{\circ}$ and decrease gradually toward the plane, reaching 0.1 at $b = 0^{\circ}$. Each frequency map was multiplied by this latitude-dependent weight, after which the maps were smoothed with the telescope beam and instrumental noise was added, following the same procedure as for the unmasked simulations. We then applied all foreground removal methods to these artificially dimmed maps using the Gaussian oscillating beam. The reconstructed spectra are shown in Fig.~\ref{fig:dimmed_results}. This mock masking improves the performance of all methods, with the most significant gains seen for PCA, FastICA, and SDecGMCA. 

In the frequency domain, the spurious peak persists for PCA and FastICA but is suppressed to below the 2\% level for SDecGMCA. This confirms the result already reported by \citet{matshawule2021}, namely that the peak is primarily driven by lines of sight dominated by exceptionally bright foregrounds. Our findings therefore strengthen the case that excluding such regions prior to component separation can substantially improve the recovery of the cosmological signal.

In the angular domain, dimming the Galactic plane improves the performance of PCA, FastICA, and SDecGMCA alike, extending the multipole range over which the HI signal is accurately reconstructed. This result suggests that masking bright regions prior to foreground removal can be beneficial not only for SDecGMCA but also for other BSS methods, particularly for the angular power spectrum.

\section{Conclusions}
\label{sect:conclusions}

We assessed the performance of SDecGMCA, an extension of DecGMCA to spherical data that jointly performs source separation, for HI 21\,cm intensity mapping under increasingly realistic beam conditions. We also compared to widely used and known foreground removal techniques, both model-fitting and BSS methods. Our main findings can be summarized as follows:

\begin{itemize}
    \item Under the simplest Gaussian degraded beam, all methods recover the HI angular and frequency power spectra with better than 10\% accuracy across a broad range of scales, as expected given the preserved spectral smoothness of the foregrounds.
    
    \item When the beam evolves smoothly with frequency, the performance of the methods diverges. Parametric fitting fails entirely, GMCA starts degrading at intermediate scales in the angular ($\ell \sim 40$) and frequency ($k_{\nu} \sim 0.2$) power spectrum, while PCA, FastICA, and SDecGMCA maintain accuracy over wider ranges in both angular ($4 \lesssim \ell \lesssim 200$) and frequency ($k_{\nu} > 0.1$) domains.
    
    \item In the most realistic and challenging oscillating beam case, the fitting approaches fail catastrophically, and only BSS methods remain viable. Among them, SDecGMCA achieves the most accurate reconstruction, suppressing the spurious peak at $k_\nu \sim 0.3$ and recovering the angular spectrum at intermediate multipoles ($20 < \ell < 200$). Nevertheless, beam inversion remains a fundamental limitation, producing unstable results beyond $\ell \sim 200$.
    
    \item The number of sources $n_s$ is a critical hyperparameter for BSS methods. We find that a moderate choice of $n_s \simeq 5$--7 provides the best balance between underfitting and overfitting, with the precise optimum being scale-dependent. This sensitivity highlights the need for robust, data-driven strategies for setting $n_s$ in real observations, where the ground truth is not available.
    
    \item We further explored masking strategies. Post-processing by excluding foreground-contaminated lines of sight substantially reduces residual artifacts in SDecGMCA, including the suppression of the spurious peak and improved recovery at low multipoles. Simulating a Galactic mask by dimming the Galactic plane demonstrated that such masking improves the performance of PCA, FastICA, and especially SDecGMCA, extending the multipole range over which the HI signal can be reliably reconstructed.
\end{itemize}

Taken together, our results highlight both the promise and the limitations of current approaches. Simple fitting methods are inadequate once realistic instrumental effects are introduced, while standard BSS techniques such as PCA, FastICA, and GMCA only partially cope with beam distortions. SDecGMCA stands out as a very promising option, consistently performing across all tested scenarios. However, its dependence on hyperparameters (particularly $n_s$ and the regularization scheme) and the intrinsic instability of beam inversion remain major challenges. 

Future work should focus on three directions: (i) incorporating masking strategies directly into the SDecGMCA framework; (ii) developing reliable diagnostics for setting hyperparameters from data; and (iii) refining the thresholding scheme (Bonjean et al., submitted). Even without these refinements, our results show that SDecGMCA already delivers robust reconstructions on full-sky simulations without pre-masking, establishing it as a particularly promising technique for upcoming 21\,cm intensity mapping surveys.

\begin{acknowledgements}
      This work was funded by the TITAN ERA Chair project (contract no. 101086741) within the Horizon Europe Framework Program of the European Commission, the French Agence Nationale de la Recherche (ANR-22-CE31-0014-01 TOSCA) and the French Programme National de Cosmologie et Galaxies (PNCG project CIMES).
\end{acknowledgements}

\bibliographystyle{aa}
\bibliography{aa}

\appendix

\section{Complementary diagnostics}
\label{ap:extra_diagnostics}

We present additional quantitative diagnostics for the case of the most challenging beam configuration, the oscillating beam, that further assess the performance of the foreground removal methods in reconstructing the cosmological HI signal. These metrics provide complementary insight beyond the power-spectrum analysis discussed in Sect.~\ref{sect:results}.

\subsection*{Pearson correlation coefficient per map}

\begin{figure}[h!]
    \centering
    \includegraphics[width=1\linewidth]{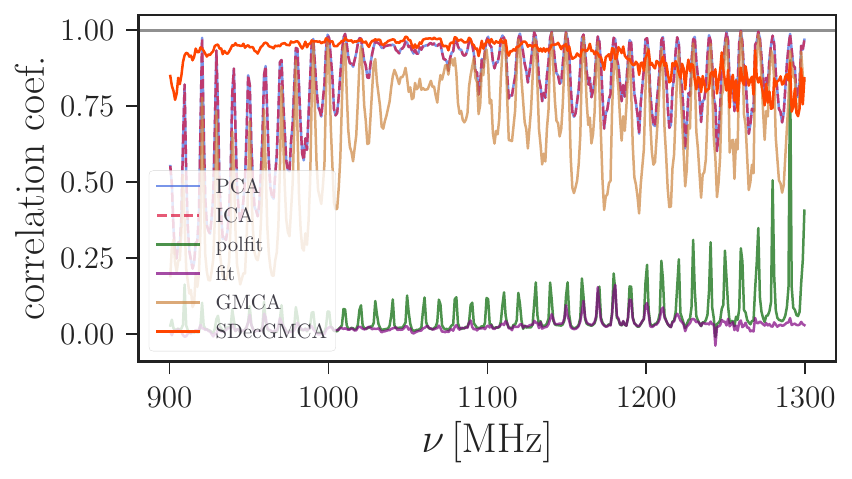}
    \caption{Pearson correlation coefficient between the reconstructed and true HI maps as a function of frequency. Values close to unity indicate an accurate recovery of the true HI structure, while lower values reveal residual foreground contamination or signal loss.}
    \label{fig:pearson_coeff}
\end{figure}

As a first additional diagnostic, we compute the Pearson correlation coefficient between the reconstructed and true HI maps at each frequency channel. This coefficient quantifies the degree of linear correlation between the spatial fluctuations of the reconstructed ($T^{\mathrm{rec}}$) and input ($T^{\mathrm{true}}$) HI brightness temperature maps:
\begin{equation}
    r(\nu) \;=\;
    \frac{
    \sum_i^{N_{\rm pix}} \left( T^{\mathrm{true}}_i(\nu) - \overline{T^{\mathrm{true}}(\nu)} \right)
             \left( T^{\mathrm{rec}}_i(\nu) - \overline{T^{\mathrm{rec}}(\nu)} \right)
    }{
    \sqrt{
    \sum_i^{N_{\rm pix}} \left( T^{\mathrm{true}}_i(\nu) - \overline{T^{\mathrm{true}}(\nu)} \right)^2
    \sum_i^{N_{\rm pix}} \left( T^{\mathrm{rec}}_i(\nu) - \overline{T^{\mathrm{rec}}(\nu)} \right)^2
    }} ,
    \label{eq:pearson_coeff}
\end{equation}
where the summation runs over all sky pixels $i$, from 1 to $N_{\rm pix}$. Values of $r(\nu)$ close to unity indicate a faithful recovery of the spatial HI structure at frequency $\nu$, whereas lower values point to residual foreground contamination or signal loss. As shown in Fig.~\ref{fig:pearson_coeff}, SDecGMCA consistently achieves the highest correlation across frequencies, demonstrating superior reconstruction fidelity compared to the other methods.

\subsection*{Variance of the fractional $C_{\ell}$ difference}

\begin{figure}[h!]
    \centering
    \includegraphics[width=1\linewidth]{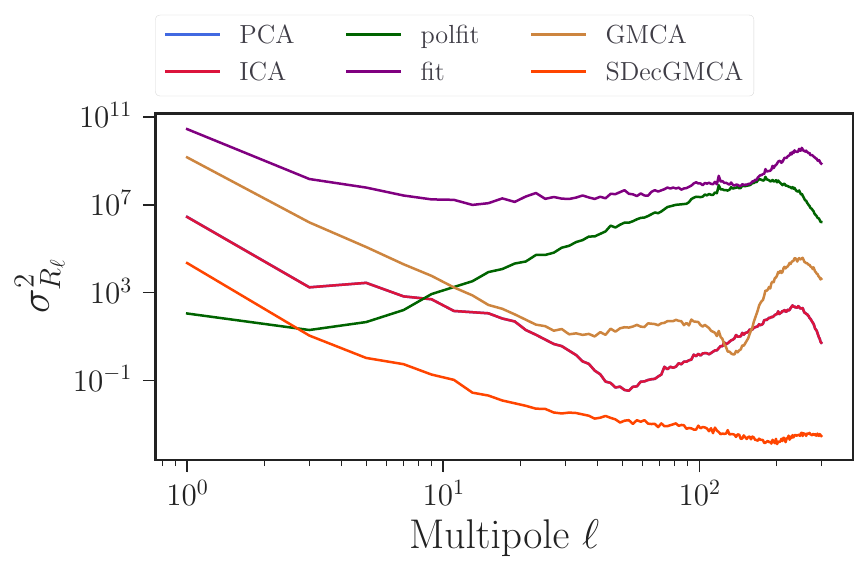}
    \caption{Variance of the fractional $C_{\ell}$ difference as a function of multipole. Lower values correspond to more uniform and consistent performance.}
    \label{fig:R_ell_variance}
\end{figure}

To complement the map-based correlation analysis, we evaluate the variance of the fractional residual difference in the angular power spectrum,
\begin{equation}
    R_{\ell} \;=\; \frac{C_{\ell}^{\mathrm{true}} - C_{\ell}^{\mathrm{rec}}}{C_{\ell}^{\mathrm{true}}},
    \label{eq:R_ell_def}
\end{equation}
averaged over frequency channels. This quantity characterizes the relative deviation of the reconstructed HI power from the true one at each multipole. We then compute the variance of $R_{\ell}$ across all frequency channels as
\begin{equation}
    \sigma^2_{R_{\ell}} \;=\;
    \left\langle \left( R_{\ell} - \langle R_{\ell} \rangle \right)^2 \right\rangle_{\mathrm{channels}} ,
    \label{eq:R_ell_var}
\end{equation}
which provides a quantitative measure of the stability and consistency of each method across scales. Lower values of $\sigma^2_{R_{\ell}}$ indicate more uniform performance and reduced scale-dependent bias. As illustrated in Fig.~\ref{fig:R_ell_variance}, SDecGMCA exhibits the smallest variance at intermediate multipoles ($10 \lesssim \ell \lesssim 200$), confirming its robustness and stability in reconstructing the HI signal.

\section{Masking in frequency space}
\label{ap:hi_bright_masking}

\begin{figure}[h!]
    \centering
    \begin{minipage}{0.48\textwidth}
        \centering
        \includegraphics[width=\textwidth]{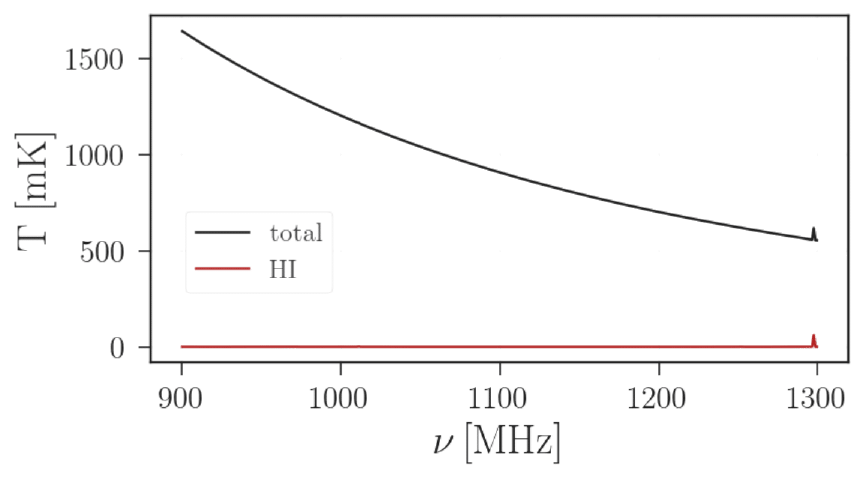}
    \end{minipage}
    \hfill
    \begin{minipage}{0.5\textwidth}
        \centering
        \includegraphics[width=\textwidth]{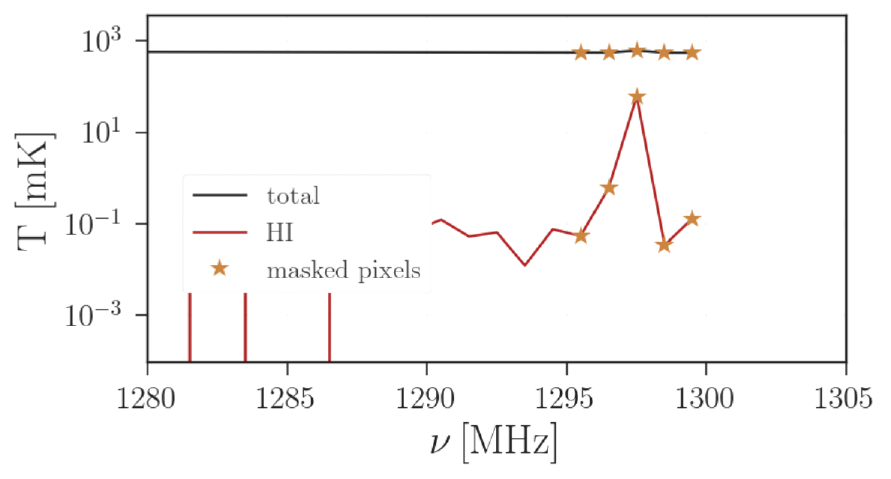}
    \end{minipage}
    \caption{Temperature profile for a single line of sight. \emph{Top:} Total observed brightness temperature $T(\nu)$ (black) and the underlying HI signal along the same LOS (red). \emph{Bottom:} Zoom on the frequency range containing the prominent peak (note the logarithmic $T$ scale). Star symbols mark the channels flagged by our sigma-clipping criterion and replaced by the second-pass polynomial baseline (Appendix~\ref{ap:hi_bright_masking}). This example corresponds to the line of sight hosting the brightest HI source in the simulation and illustrates the masking procedure.}
    \label{fig:los_with_peak}
\end{figure}

\begin{figure}[h!]
    \centering
        \includegraphics[width=\linewidth]{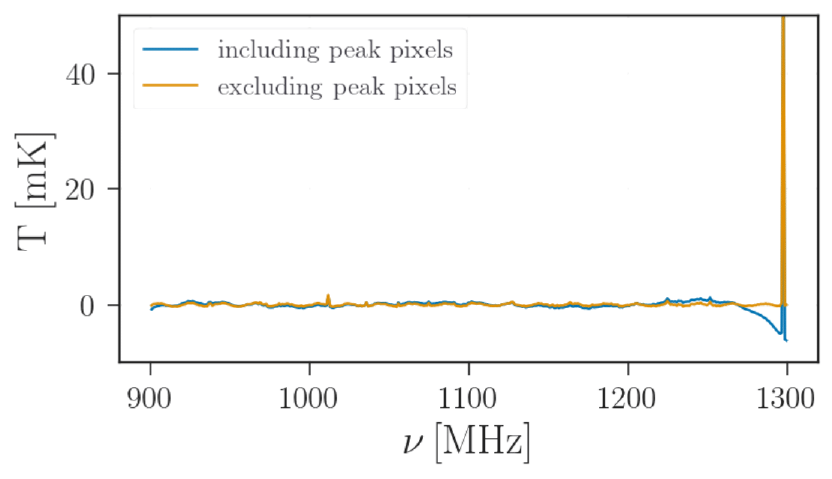}
    \caption{Flattened signal after baseline subtraction. The blue curve shows the result when all pixels are included in the polynomial fitting, while the orange curve shows the improved baseline obtained when bright outlier pixels are masked and excluded from the second fit.}
    \label{fig:los_with_peak_flattened}
\end{figure}

In this appendix, we describe in detail the pre-processing step used to identify and mask sharp peaks in the temperature profiles, which can bias foreground removal. The procedure begins by fitting a sixth-order polynomial along each line of sight (LOS) in the total observed temperature maps (see top panel of Fig.~\ref{fig:los_with_peak}). Subtracting this polynomial from the original data flattens the spectra, as illustrated in Fig.~\ref{fig:los_with_peak_flattened}. To identify and exclude bright peaks, we apply sigma clipping with a threshold of $5\sigma$ to the flattened residuals. Pixels flagged as outliers (stars in bottom panel of Fig.~\ref{fig:los_with_peak}) are masked. 

A second polynomial fit is then performed on the original total temperature profile, this time excluding the masked pixels. Subtracting this refined baseline yields a more accurate estimate and avoids bias from outliers. Fig.~\ref{fig:los_with_peak_flattened} compares the results of baseline subtraction with and without masking. The blue curve shows the flattened signal when all pixels are included, while the orange curve demonstrates the improved baseline removal obtained after excluding bright outliers.

\begin{figure}[h!]
    \centering
        \includegraphics[width=\linewidth]{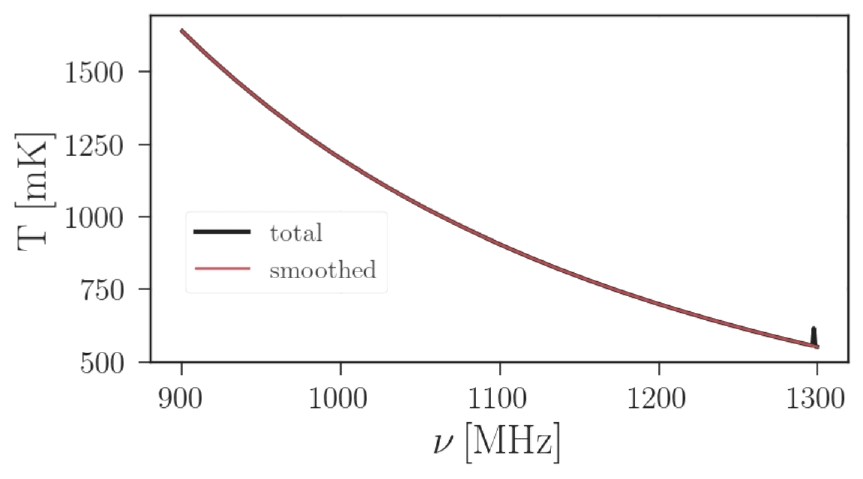}
    \caption{Temperature profile along a single LOS. The black line shows the original total signal, while the red line shows the smoothed signal obtained after applying the masking procedure described in Appendix~\ref{ap:hi_bright_masking}.}
    \label{fig:smoothed_los}
\end{figure}

Finally, in the pre-processing strategy used throughout Sect.~\ref{subsect:HI_bright_masking}, we replaced the masked pixel values with the corresponding baseline values. This smoothing procedure is illustrated in Fig.~\ref{fig:smoothed_los}, where the black line shows the original total signal and the red line shows the smoothed result. This masking was applied consistently to all input temperature maps prior to running either the fitting-based or BSS-based foreground-removal methods, and it substantially improved the identification of foreground-dominated channels, leading to a more accurate HI reconstruction.

\end{document}